\documentclass[a4paper,12pt]{article}
\pdfoutput=1
\usepackage{amssymb,amsmath,bm}
\usepackage{a4wide}
\usepackage{color}
\usepackage{slashed}
\usepackage{cite}
\usepackage{hyperref}
\usepackage{units}
\usepackage{relsize}
\usepackage{bbold}
\usepackage[latin1]{inputenc}
\usepackage{amsfonts}
\usepackage{lscape}
\usepackage{amsthm}
\usepackage{booktabs}
\usepackage{array}
\usepackage{rotating}
\usepackage{float}
\usepackage{multirow}
\usepackage{xfrac}
\usepackage{graphicx,rotating,amsmath,multirow,xcolor,colortbl,xcolor}
\usepackage{hyperref,pdflscape,slashed}
\usepackage{easyReview}
\usepackage{cleveref}

\usepackage{tikz}
\usetikzlibrary{matrix,decorations.pathreplacing}

\usepackage{cite}

\definecolor{rosso}{cmyk}{0,1,1,0.4}
\definecolor{rossos}{cmyk}{0,1,1,0.55}
\definecolor{rossoc}{cmyk}{0,1,1,0.2}
\definecolor{blu}{cmyk}{1,1,0,0.3}
\definecolor{blus}{cmyk}{1,1,0,0.6}
\definecolor{bluc}{cmyk}{1,1,0,0.1}
\definecolor{verde}{cmyk}{0.92,0,0.59,0.25}
\definecolor{verdec}{cmyk}{0.92,0,0.59,0.15}
\definecolor{verdes}{cmyk}{0.92,0,0.59,0.4}
\definecolor{Gray}{gray}{0.95}

\font\tenrsfs=rsfs10 at 12pt
\font\sevenrsfs=rsfs7
\font\fiversfs=rsfs5
\newfam\rsfsfam
\textfont\rsfsfam=\tenrsfs
\scriptfont\rsfsfam=\sevenrsfs
\scriptscriptfont\rsfsfam=\fiversfs
\def\mathscr#1{{\fam\rsfsfam\relax#1}}

\usepackage[small,bf]{caption}
\hypersetup{colorlinks,bookmarksopen,bookmarksnumbered,
linkcolor=blus,pdfstartview=FitH,urlcolor=blue,citecolor=verde}

\newcommand{\tev}{{\rm TeV}}

\newcommand{\mgfive}{MG5\_aMC@NLO}

\newcommand{\lsim}{\stackrel{<}{_\sim}}
\newcommand{\gsim}{\stackrel{>}{_\sim}}

\newcommand{\be}{\begin{equation}}
\newcommand{\ee}{\end{equation}}
\newcommand{\bea}{\begin{eqnarray}}
\newcommand{\eea}{\end{eqnarray}}
\newcommand{\beq}{\begin{equation}}
\newcommand{\eeq}{\end{equation}}
\newcommand{\beqa}{\begin{eqnarray}}
\newcommand{\eeqa}{\end{eqnarray}}

\newcommand{\sigmatot}{\sigma_{\sq\sq+\sq\sq^{*}}}
\newcommand{\nnllfast}{NNLO$_{\rm approx}$+NNLL}

\def\app#1{Appendix~\ref{#1}}
\def\eq#1{Eq.~(\ref{#1})}
\def\fig#1{Fig.~\ref{#1}}

\def\tab#1{Table~\ref{#1}}

\counterwithin*{equation}{section}

\def\eq#1{Eq.~(\ref{#1})}
\def\fig#1{Fig.~\ref{#1}}
\def\sec#1{Sec.~\ref{#1}}
\def\app#1{Appendix~\ref{#1}}

\def\lsp{\tilde\chi_1^0}
\def\mlsp{m_{\tilde\chi_1^0}}
\def\rap{r_{\tilde g \tilde q}}
\def\rapW{r_{\tilde W \tilde q}}
\def\rapB{r_{\tilde B \tilde q}}

\def\rapinv{r_{\tilde q \tilde g}}   %

\newcommand{\sq}{\tilde q}
\newcommand{\gl}{\tilde g}
\newcommand{\msq}{m_{\tilde q}}
\newcommand{\mgl}{m_{\tilde g}}
\newcommand{\shat}{\hat s}
\newcommand{\fb}{~\mathrm{fb}}
\newcommand{\ab}{~\mathrm{ab}}
\newcommand{\TeV}{~\mathrm{TeV}}
\newcommand{\GeV}{~\mathrm{GeV}}

\definecolor{nicegreen}{rgb}{0.1,0.5,0.1}
\definecolor{nicered}{rgb}{0.5,0.1,0.1}
\definecolor{violet}{rgb}{0.7,0.3,0.3}

\begin{document}

{\hfill CERN-TH-2026-240}

{\hfill UMD-PP-026-08}

\vspace{2cm}

\begin{center}
\boldmath

{\textbf{\LARGE Shortcomings of the squark simplified model}}

\unboldmath

\bigskip

\vspace{1 truecm}

{\bf Kaustubh Agashe}$^a$, {\bf Roberto Franceschini}$^{b,d}$,\\{\bf Gian F. Giudice}$^{c,d}$, {\bf Andre Lessa}$^e$, {\bf Xing Wang}$^{b,f}$
 \\[5mm]

$^a$ {\it Maryland Center for Fundamental Physics, University of Maryland, \\ College Park, MD 20742, USA}\\[2mm]
$^b$ {\it Universit\`a degli Studi Roma Tre and INFN, 00146 Rome, Italy}\\[2mm]
$^c$ {\it New York University Abu Dhabi, UAE}\\[2mm]
$^d$ {\it CERN, Theoretical Physics Department, Geneva, Switzerland}\\[2mm]
$^e$ {\it Instituto de F\'isica, Universidade de S\~ao Paulo, 05315-970 S\~ao Paulo, Brazil}\\[2mm]
$^f$ {\it Institut de F\'isica d'Altes Energies (IFAE), \\ The Barcelona Institute of Science and Technology (BIST), \\ Campus UAB, 08193 Bellaterra, Barcelona, Spain}\\[2mm]

\vspace{2cm}

{\bf Abstract }
\end{center}

\begin{quote}
We argue that the squark simplified model commonly used in collider analyses fails to represent realistic low-energy supersymmetric models and significantly underestimates squark production rates at hadron colliders. This has important implications for extracting squark mass limits at the LHC and assessing the potential reach of HL-LHC or the FCC-hh. In particular, we find that the current LHC squark mass limit is realistically in the range of 2.6--3.0 TeV, rather than the 1.7--1.8 TeV range that is typically quoted based on the simplified model. 
\end{quote}

\thispagestyle{empty}
\vfill

\newpage
\tableofcontents

\section{Introduction}
\label{sec:intro}
Simplified models are valuable tools for capturing the essential features of complex  theories with many free parameters. By focusing on just a few degrees of freedom and parameters, these models can provide approximate yet practical descriptions of individual elements of the full theory. A typical application of simplified models is found in collider analyses of low-energy supersymmetry, where the wide array of experimental signals---characterized by numerous new particles and unknown parameters---is streamlined into more manageable structures, often considering only one or two particles at a time.

However, as crisply expressed by a remark often attributed to Einstein, ``Everything should be made as simple as possible, but not simpler"~\cite{Einstein}. The purpose of this paper is to demonstrate that the squark simplified model commonly used in LHC analyses makes things simpler than they actually are.

The squark simplified model consists of eight degenerate color-triplet complex scalars with a 100\% decay mode $\sq \to q \lsp$, where $\lsp$ is an invisible neutralino. It is assumed that no other supersymmetric particle plays any significant role in squark production, decay processes, and the kinematics of the produced squarks and their decays.  The model is defined by only two parameters: the squark mass $\msq$ and the mass of the lightest neutralino $\mlsp$. Because of the absence of other new particles, the squark production at high-energy colliders is  fully determined by QCD interactions and is a function of $\msq$ alone. The kinematics of the decay process, which characterizes the experimental signatures, is governed by the mass difference  $\msq -\mlsp$.

At first sight, it may seem that the squark simplified model catches the essential properties of squarks in a vast class of supersymmetric theories. On the contrary, we will argue that this is not the case and that the squark simplified model fails to represent realistic models of low energy supersymmetry.

The central observation of our paper is that the gluino cannot be decoupled from the squark simplified model without conflicting with a basic theoretical principle. In the limit of heavy gluinos, squarks are effectively described by a non-supersymmetric theory. Hence, their masses quickly become vulnerable to large quantum corrections, which are quadratically sensitive to the UV cutoff. As a result, any hierarchy in the gluino/squark mass ratio $\rap =\mgl /\msq$ is rapidly destabilized at the quantum level. 
We will demonstrate that, under any realistic circumstances, supersymmetric models cannot support values of $\rap$ larger than about 2--3. Since the squark simplified model corresponds to $\rap \to \infty$, it cannot reliably represent realistic models of low-energy supersymmetry. 

The issue is not merely theoretical, but has practical implications for experimental analyses at the LHC and future colliders. This is because the squark production cross section at hadron colliders receives large contributions from $t$-channel gluino exchange. For instance, for $\msq = 3 \TeV$ and $\rap=2$, the total squark production rate at the LHC is more than 60 times larger than what predicted by the simplified model. Taking into account the upper bound $\rap \lsim 2$--3, it is clear that ignoring the gluino severely underestimates the experimental constraints on squarks. Indeed, the PDG~\cite{ParticleDataGroup:2024cfk} and reviews on LHC data quote $\msq >$ 1.7--1.8 TeV as the current mass limit for the first two generations of squarks, based on simplified models. On the contrary, by focusing on the relevant region $1<\rap \lsim 2$--3,  we show that the current limit on squark masses in any realistic situation is in the range 2.6--3.0 TeV.

We emphasize that presenting experimental results in the limit $\rap \to \infty$ does not offer a more conservative approach, but only misrepresents the predictions of low-energy supersymmetric models.
Correcting this misrepresentation also clarifies the relative constraining power of colored and electroweak sparticle searches in realistic supersymmetric models.

Another important result of our analysis is that the gluino contribution to squark production becomes more prominent as the luminosity increases and, therefore, its importance grows as we will enter the HL-LHC phase. Moreover, our considerations apply also to future hadron colliders, such as the FCC-hh, significantly improving their squark discovery reach.

Our conclusion is that the squark simplified model is inadequate to represent the wide range of realistic supersymmetric models and gives misleading results when determining bounds on the squark mass. 
However, we will show that these limitations can be addressed by using an alternative parametrization that captures the core features of realistic supersymmetric models and provides reliable results for collider searches.

The paper is organized as follows. In \sec{sec:ratio}, we discuss the origin of the upper bound on $\rap$ in realistic supersymmetric models. Section~\ref{sec:anatomy} provides a detailed analysis of the structure of the squark production cross section and its dependence on $\rap$. In \sec{sec:reach}, we quantify how our findings affect the current LHC mass limit on squark masses as well as the potential reach of future collider projects, such as the HL-LHC and FCC-hh. Section~\ref{sec:ew-inos} examines the impact of virtual exchange of electroweak gauginos on the squark production cross section, considering both realistic models and the simplified model. In \sec{sec:gluino} we investigate the validity of the simplified model for gluino production. Finally, \sec{sec:conc} presents our conclusions. Some technical details on how we extrapolated the ATLAS squark analysis to our case are contained in the appendices. 

\section{The upper bound on the gluino/squark mass ratio}
\label{sec:ratio}

In a generic supersymmetric model, gluino and squark masses are independent parameters. At first glance, this may suggest that their ratio $\rap =\mgl /\msq$ could take any arbitrary value. However, we will argue that this is not the case. Realistic supersymmetric models cannot achieve a ratio $\rap$ greater than a few, due to a fundamental property of quantum field theories.

The key observation of this paper is that, in the presence of a scale separation between the gluino and squark masses, there is an energy range ($\msq < E < \mgl$) in which the effective theory describing squarks feels hard supersymmetry breaking. In this case, squarks, being generic spin-0 particles not protected by supersymmetry or non-linearly realized global symmetries, must receive contributions to their masses which, at the quantum level, become quadratically sensitive to the energy cutoff of the effective theory. This is exactly analogous to the naturalness problem encountered with the Higgs mass. The difference with the SM is that here the problem is even more tangible because we know the completion of the theory above the energy cutoff. In the case of squark masses,
 the ``quadratic divergent" quantum corrections are calculable and proportional to $\mgl^2$. This means that large values of  $\rap$ are quantum mechanically unstable and rapidly driven to values of order one.

Note that the quantum instability does not occur for large values of $1/\rap=\msq/\mgl$ which, unlike the case of large $\rap$, describe perfectly plausible situations. This is because $\mgl$ carries an $R$-charge and, for symmetry reasons, cannot be renormalized by the $R$-neutral squark mass alone. Indeed, the case of very large $1/\rap$ is successfully realized in Split Supersymmetry~\cite{Arkani-Hamed:2004ymt,Giudice:2004tc,Arkani-Hamed:2004zhs}.

The sensitivity of the squark mass to $\mgl$ can be easily quantified. With an RG resummation of the QCD logs at leading order, we find that $\mgl$ quantum-mechanically feeds an additive contribution to the squark square mass
\beq
\Delta \msq^2 =\frac{8}{9} \left[ 1-\left( 1+\frac{3\alpha_s(\mgl)}{2\pi}\, \ln \frac{M}{\mgl}\right)^{-2}\right]  \mgl^2 \, ,
\label{susy1}
\eeq
where $M$ is the scale at which the gluino mass is  generated. For example, for $\mgl =3$~TeV (but the dependence of \eq{susy1} on $\mgl$ is only logarithmic), we find  
\beq
\rap =\frac{\mgl}{\sqrt{\Delta \msq^2}}=\left\{
\begin{array}{cl}
1.2 & {\textrm{for}}~M=10^{16}\GeV \\[2mm]
2.2 & {\textrm{for}}~M=10^{5}\GeV
\end{array}
\right.
\eeq
We will later present a more accurate NLO calculation, but this LO estimate is sufficient to understand the effect.

The gluino contribution is very efficient in feeding mass to squarks and destabilizing any large value of $\rap$. As $M$ is lowered, $\rap$ grows, but remains of order unity for any realistic value of the scale at which supersymmetry breaking is communicated to the visible sector.

To establish a realistic upper bound on $\rap$, we make the extreme hypothesis that the entire squark mass is due to gluino quantum effects. This corresponds to the condition \beq
\rap (M)=\infty \, ,
\label{boundary}
\eeq
where $M$ is the scale of supersymmetry-breaking mediation. This procedure corresponds to taking vanishing squark masses at the scale $M$ and is intended to maximize the value of $\rap$. Note that the boundary condition in \eq{boundary} is not merely academic, but applies to realistic models, such as gaugino mediation~\cite{Kaplan:1999ac,Chacko:1999mi} or conformal sequestering~\cite{Luty:2001jh,Luty:2001zv,Perez:2008ng}. However, our goal is not to study any specific model, but to derive a reliable theoretical upper bound on $\rap$.

\begin{figure}[htb]
\centering
\includegraphics[width=0.80\textwidth]{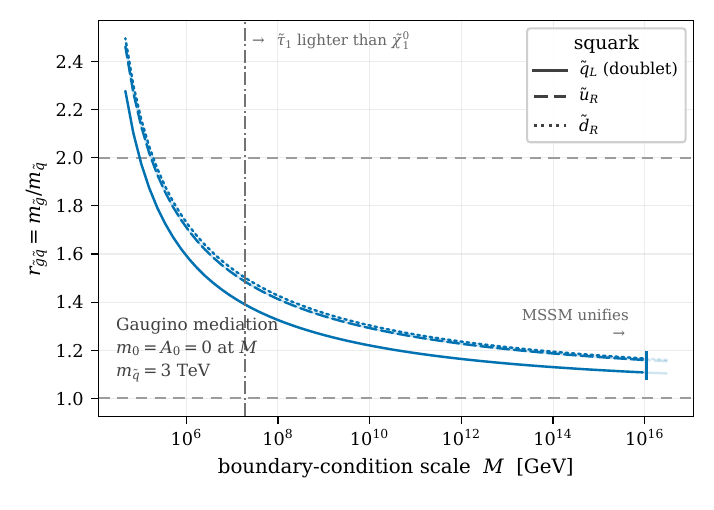}
\caption{The ratio $\rap$ using the boundary condition in \eq{boundary}, shown as a function
of $M$ at fixed light-squark mass $3\TeV$, resolved by squark species. Gaugino masses
are unified at $M$, with $m_0=A_0=0$, $\tan\beta=10$ and $\mu>0$. The unified gaugino mass is determined such that, for at each $M$, the squark pole mass (averaged over quark species) is $3\TeV$. Following this procedure, the unified gaugino mass varies from $6.7$ to $1.6\TeV$ across the plot. Horizontal dashed
lines mark $\rap=1$ and $2$. To the right of the dash-dotted line the lighter stau is
the lightest supersymmetric particle, in conflict with the simplified-model assumption that $\sq\to q\lsp$ is the dominant decay mode. In the region to the left of the vertical line the lightest state is a neutralino, but
higgsino-like, with a mass that varies between $140$ and $870\GeV$. The stau is lighter than the Bino for any value of $M$.}
\label{fig:gaugino}
\end{figure}

The result of NLO RG evolution starting from \eq{boundary} is shown in \fig{fig:gaugino}. Typically, the ratio $\rap$ ranges between 1 and 2. Even if we assume that the dynamics of supersymmetry breaking and supersymmetry mediation occur at a scale $M$ as low as 50~TeV, $\rap$ is at most equal to 2.5. It is not easy to imagine realistic scenarios in which $M$ is much lower than 50~TeV. This leads us to the expectation that $\rap$ cannot reasonably exceed about 2.5.

The prediction for $\rap$ shown in \fig{fig:gaugino} is based on the extreme hypothesis of \eq{boundary}, which corresponds to the absence of squark soft masses at the mediation scale. However, smaller values of $\rap$ are generically obtained in models in which squark and gluino masses are simultaneously generated by the mechanism of communicating supersymmetry breaking.

A typical example of a supersymmetric model in which both $\msq$ and $\mgl$ are non-vanishing is gauge mediation~\cite{Dine:1993yw,Dine:1994vc,Dine:1995ag,Giudice:1998bp}. In this case, $\rap$ is calculable in terms of the two parameters of the model: the messenger mass $M_{\textrm{mess}}$ and the messenger index $N$. The boundary condition for $\rap$ at the messenger mass is
\beq
\rap (M_{\textrm{mess}})= \sqrt{\frac{3\, N}{8\, Z}}  \, , ~~~ Z=\left\{
\begin{array}{ll}
1+\frac{9\,\alpha_2^2}{16\, \alpha_3^2}+\frac{\alpha_1^2}{80\, \alpha_3^2} & {\textrm{for}}~{\tilde q}_L \\
1+\frac{\alpha_1^2}{5\, \alpha_3^2} & {\textrm{for}}~{\tilde u}_R \\
1+\frac{\alpha_1^2}{20\, \alpha_3^2} & {\textrm{for}}~{\tilde d}_R
\end{array} \right.
\label{gmedb}
\eeq
where $\alpha_{1,2,3}$ describe the SM gauge couplings evaluated at the scale $M_{\textrm{mess}}$ and with unified normalization.

\begin{figure}[htb]
\centering
\includegraphics[width=0.80\textwidth]{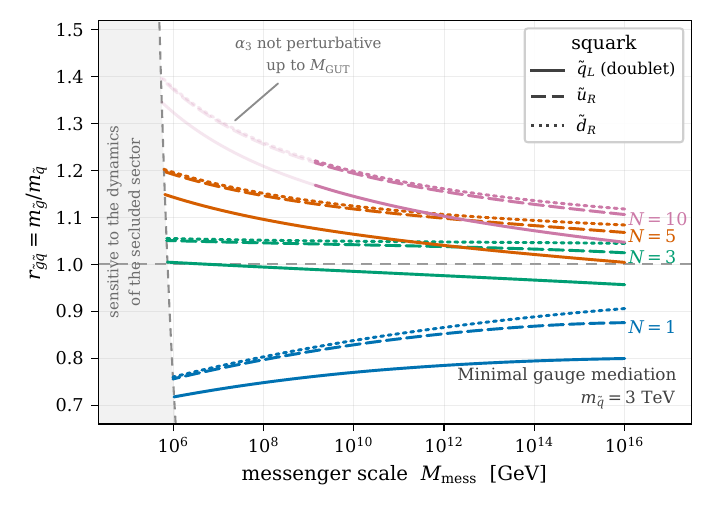}
\caption{The ratio $\rap$ in gauge mediation versus the messenger mass, at fixed
light-squark mass $3\TeV$, resolved by squark species (line style) and messenger index $N$. The shaded region is excluded by the theoretical requirement of avoiding sensitivity to the (unknown) dynamics of the secluded sector. Lines with lighter shades correspond to regions where
$\alpha_{s}$ becomes non-perturbative before $10^{16}$ GeV (only $N=10$ is
affected).}
\label{fig:gmsb}
\end{figure}

By evolving the condition in \eq{gmedb} to low energies, we find the values of $\rap$ shown in \fig{fig:gmsb}. In our calculation, the overall mass scale has been fixed by taking $\msq =$ 3 TeV, which is in the interesting ballpark for our considerations. As $M_{\textrm{mess}}$ becomes too low, one reaches a situation in which there is no longer a separation between $M_{\textrm{mess}}$ and the scale of supersymmetry breaking. In this region, the mass spectrum becomes sensitive to the (unknown) details of the secluded sector and the perturbative calculation goes out of control. In \fig{fig:gmsb}, we have excluded this region by estimating the values of $M_{\textrm{mess}}$ for which the messenger threshold corrections~\cite{Dimopoulos:1996gy} become sizable. For large $N$ and low $M_{\textrm{mess}}$, the QCD gauge coupling enters a non-perturbative regime before $10^{16}$ GeV, thus preventing a successful perturbative gauge coupling unification. In \fig{fig:gmsb}, we have used lighter shading to indicate when this occurs.

As shown in \fig{fig:gmsb}, gauge mediation predicts values of $\rap$ that can only be moderately larger than 1, and clearly smaller than the extreme case studied before, in which $\rap =\infty$ at the matching scale. The gauge-mediation boundary condition for $\rap$ grows with the messenger index $N$, as shown by \eq{gmedb}. However, the effect is moderated by the RG evolution, as apparent from \fig{fig:gmsb}. For low $M_{\textrm{mess}}$, the energy range for RG running is smaller, and the RG dampening of $\rap$ is reduced. Nevertheless,
even taking the extreme case $N=10$ and $M_{\textrm{mess}}$ as low as to renounce to perturbative unification, $\rap$ is at most 1.4. The message from \fig{fig:gmsb} is that, as far as gauge mediation is concerned, $\rap$ can never be much larger than one.

The goal of this paper is to identify the parameter range of a simplified model that can be representative of low-energy supersymmetry, while respecting the properties of quantum field theory. Our results show that introducing the parameter $\rap$ is necessary, but not all choices lead to realistic models. We have found that the gluino/squark mass ratio is bounded from above, with $\rap \lsim 2$--3. The simplified model currently used in collider analyses assumes $\rap =\infty$ and clearly violates our upper bound.

One could imagine to weaken the bound on $\rap$ by choosing tachyonic boundary conditions for squark masses at a high-energy scale $M$, adjusted in such a way that the physical values of $\msq^2$ are positive, but nearly vanishing. This requires an extreme fine tuning between the gluino mass and the tachyonic boundary conditions for each squark, stop included. Moreover, tachyonic boundary conditions would invalidate the mechanism of radiative EW breaking. Finally, they would generate a vast pattern of vacua that are deeper than the usual EW Higgs vacuum, spontaneously breaking color and electromagnetism. Whether these vacua are lethal for our universe depends on tunneling rates and the cosmological evolution of scalar fields. All together, tachyonic boundary conditions seem artificial, not physically motivated, and lacking any realistic theoretical realization, but not mathematically impossible.

However, one must keep in mind the goal of a simplified model, which is to capture the essential features of the underlying theories. Any squark mass limit derived under the condition $\rap \gsim 3$ must be accompanied by the warning that the limit is valid only for hypothetical models with tachyonic tuning. Evidently, mass limits valid only under such pathological assumptions are not particularly informative. Our concern is to derive limits that apply to generic conditions of realistic supersymmetric models. Considering $\rap \gsim 3$ is {\it not more conservative}, but only {\it less representative} of low-energy supersymmetric models. The choice $\rap = \infty$ is the extreme example.

A more interesting exception can be found beyond the ordinary scenarios with $\mathcal{N}=1$ supersymmetry. This is the case of supersoft models based on $\mathcal{N}=2$ supersymmetry in the gauge sector and $\mathcal{N}=1$ in the matter sector~\cite{Fox:2002bu}. Gauginos acquire Dirac masses by mixing with new chiral superfields in the adjoint gauge representations and squark masses receive only finite quantum corrections, leading to
\beq
\rap=\sqrt{\frac{3\, \pi}{4\alpha_s\, \ln(m_a^2/\mgl^2)}}\, .
\eeq
Here $m_a$ is the mass of the real scalar component of the new adjoint chiral superfield. Depending on the value of $m_a$, the ratio $\rap$ could now become (mildly) larger than 3, although the existence of a strict upper bound on $\rap$ holds true.

However, theories with partial $\mathcal{N}=2$ supersymmetry are a significant departure from the set of theories targeted by standard LHC analyses. 
Dirac gauginos with $\mathcal{N}=2$ supersymmetry lead to a rather distinct gluino and squark collider phenomenology, both in terms of production cross sections and decay modes. In particular, the production rate for squarks is suppressed, while gluino production is enhanced~\cite{Kribs:2012gx,Chalons:2018gez}. As we will show in the following, the Majorana nature of gluinos is an essential feature of our analysis. Therefore, our considerations are restricted to $\mathcal{N}=1$ supersymmetric models with Majorana gauginos.

In conclusion, because of the well-known phenomenon of quantum instability of scalar masses, a value of $\rap$ larger than about 2--3 is untenable because it would rapidly collapse to an order-one number.
This result is not based on a specific model-dependent choice, but is a general feature of a large class of theories based on low-energy $\mathcal{N}=1$ supersymmetry. Our result is a robust consequence of quantum field theory, related to the issue of naturalness of scalar masses. It follows from a well-defined calculation of quantum corrections and does not rely on any arbitrary criterion.

\section{Anatomy of the squark production cross section}
\label{sec:anatomy}

In this section, we review the various contributions to the squark production cross section at hadron colliders, focusing on the impact of the gluino-squark mass ratio $\rap$. We consider supersymmetric theories with eight degenerate squarks, and assume that the squark left-right mixing is negligible. The squark production cross section can be dissected into four channels, identified by the four combinations of $\sq\sq^*$/$\sq\sq$ and opposite/same chiralities of the squarks in the final state, as shown in \tab{tab:cros}.

To understand the behavior of the different channels, we recall that the production rate is dominated by parton collisions with energies just above the production threshold $\hat{s} =4\, \msq^2$. Therefore, information about the cross section can be derived from a partial-wave decomposition in the non-relativistic limit.

The final state in squark pair-production consists of two scalar particles, so its angular momentum is purely
orbital, $J=L$, and waves in parton collisions open up as $\hat\sigma\sim\beta^{2L+1}$ with
$\beta=\sqrt{1-4\msq^{2}/\shat}$. Hence, the leading contribution near threshold comes from the lowest allowed $L$.

There are three kinds of contributions to squark pair production. The first one comes from diagrams with a gluon $s$-channel exchange $gg,\, q \bar q\to g^* \to \sq\, \sq^*$. The spin-1 nature of the intermediate gluon forces the final squarks to be in a $J=1$ state, and therefore particle production occurs in P-wave, suppressed by $\beta^3$.

The second contribution comes from $gg$-initiated processes with $\sq$ exchange in $t/u$-channel or pointlike gluon-squark interactions (seagull diagram). Since the initial two gluons can be in a $J=0$ state, production in S-wave is allowed.

The third kind of contributions comes from $t$-channel gluino exchange.
The gluino propagator leads to two different effects:
\begin{equation}
\frac{\slashed p+\mgl}{t-\mgl^{2}}\,:\qquad
\begin{aligned}
\slashed p &\quad\text{(no chirality flip)} &&\to\ LL,\,RR, && J_{z}=\pm1,\\
\mgl &\quad\text{(Majorana chirality flip)} &&\to\ LR, && J_{z}=0.
\end{aligned}
\label{eq:numerator}
\end{equation}
The chirality-conserving ($\slashed p$) term annihilates
$q\bar q$ in opposite helicities, $J_{z}=\pm1$, which forbids $L=0$ and forces a P-wave suppression. Same-helicity $q\bar q$ states give $J_{z}=0$ and can annihilate through a $\mgl$ insertion into squarks in S-wave. Since the two terms in the gluino propagators also govern the chiralities of the colliding quarks (which are then transferred to the outgoing squarks), each squark production channel has a well-defined threshold behavior, which is shown in \tab{tab:cros}. These considerations are purely based on kinematics, the scalar nature of squarks and the Majorana nature of the gluino. Thus, they are independent of the particular scattering matrix element.

\renewcommand{\arraystretch}{1.5}
\begin{table}[t]
\centering
\begin{tabular}{|c|c|c||c|}
\hline
 {\small Parton } &  {\small Parton }&  {\small Threshold} & {\small $pp$ } \\[-3mm]
  {\small  process} &  {\small cross section}&  {\small behavior} & {\small cross section} \\
\hline
\hline
{\normalsize $gg\to \sq_{L,R}\, \sq_{L,R}^*$} & 
{\normalsize $\frac{n_f\,7\pi\alpha_{s}^{2}\,\beta}{192\,\msq^2}\,f_{\sq}(\beta)$} &
{\normalsize S-wave}& {\normalsize ${\sigma}_{0g}$}\\
& {\scriptsize $f_{\sq} (\beta)=(1-\beta^2)$}{\footnotesize $\big( \frac{41-31\beta^2}{7}-\frac{17-18\beta^2+\beta^4}{7\,\beta}\, \ln \frac{1+\beta}{1-\beta}\big)$} & & \\
&{\footnotesize $=1-\frac{40\,\beta^2}{21}+\,$}{\scriptsize $\mathcal{O}(\beta^4)$} & & \\[1mm]
\hline
{\normalsize $q_i\bar q_i\to \sq_{L,R}\,\sq_{L,R}^*$} & 
{\normalsize $\frac{n_f\,\pi\alpha_{s}^{2}\,\beta^3(1-\beta^2)}{27\,\msq^2}$} &
P-wave & {\normalsize ${\sigma}_{0q}$}
\\[1mm]
\hline
{\normalsize $q_i\bar q_j\to \sq_{iL,R}\,\sq_{jL,R}^*$} & 
{\footnotesize $\frac{4\pi\alpha_{s}^{2}\,\beta^3\,\delta_{ij}}{81\,\msq^2}/\rap^2+\frac{4\pi\alpha_{s}^{2}\,\beta^3\,(6-\delta_{ij}(1+\beta^2))}{81\,\msq^2\,(1-\beta^2)}/\rap^4$} &
P-wave & {\normalsize ${\sigma}_{1}/\rap^2+{\sigma}_{2}/\rap^4$}
\\[1mm]
\hline
\hline
{\normalsize $q_i\bar q_j\to \sq_{iL,R}\,\sq_{jR,L}^*$} & 
{\normalsize $\frac{4\pi\alpha_{s}^{2}\,\beta}{9\,\msq^2}/\rap^2$} &
S-wave & {\normalsize ${\sigma}_{3}/\rap^2$}
\\[1mm]
\hline
\hline
{\normalsize $q_i q_j\to \sq_{iL,R}\,\sq_{jL,R}$} & 
{\normalsize $\frac{4\pi\alpha_{s}^{2}\,\beta\,(3-\delta_{ij})}{27\,\msq^2}/\rap^2$} &
S-wave & {\normalsize ${\sigma}_{4}/\rap^2$}
\\[1mm]
\hline
\hline
{\normalsize $q_i q_j\to \sq_{iL,R}\,\sq_{jR,L}$} & 
{\normalsize $\frac{8\pi\alpha_{s}^{2}\,\beta^3}{27\,\msq^2\,(1-\beta^2)}/\rap^4$} &
P-wave & {\normalsize ${\sigma}_{5}/\rap^4$}
\\[2mm]
\hline
\end{tabular}
\renewcommand{\arraystretch}{1}
\caption{The different squark production channels, classified by the four possible final states. The number of squark flavors is denoted by $n_f$ and taken to be equal to 4. The gluino-independent
``simplified" cross section is given by $\sigma_{0}=\sigma_{0g}+\sigma_{0q}$, and
$\sigma_{1\ldots5}$ are the gluino-mediated contributions discarded by the simplified model. In the large-$\rap$ limit, $\sigma_{1\ldots5}$ are independent of $\rap$. The third column gives the threshold
behavior of each channel. For the gluino-mediated channels only the leading term of the expansion in $1/\rap$ is kept, except for the $q\bar q$-initiated same-chirality  channel, for which also the $1/\rap^4$ term is given because, as explained in the text, its contribution can be sizable. The indices $i,j$ label quark and squark flavors.}
\label{tab:cros}
\end{table}

\boldmath
\subsection{The large-$\rap$ behavior}
\unboldmath
The main point of our paper is the result presented in \sec{sec:ratio}, which shows that the landscape of realistic supersymmetric models populates only the region of low $\rap$, specifically up to about $\rap \approx 2$--3. Despite our finding, we begin our analysis by examining the squark production cross section in the opposite limit: large $\rap$. There are four reasons for taking this approach: {\it (i)} it will plainly reveal why and how the squark simplified model fails to describe realistic supersymmetric setups; {\it (ii)} it will allow us to factorize the cross-section dependence on $\msq$ and $\rap$, making the analysis more transparent; {\it (iii)} it will give us insights about the properties of the squark production cross section, which will be useful when we turn to the realistic region $\rap \lsim 3$; {\it (iv)} it will help us to match the cross section between the small- and large-$\rap$ regions, since the asymptotic behavior gives a reasonable approximation of the cross section even at $\rap$ as low as 3.

Using a power expansion in $1/\rap$, the squark production cross sections in the four chirality channels can be parametrized as
\bea
&&  \sigma(pp\to\sq_{L,R}\,\sq_{L,R}^*)=\sigma_0 +\sigma_1/\rap^2 +\sigma_2/\rap^4 +\mathcal{O}(1/\rap^6) \, , ~~~~\sigma_0\equiv \sigma_{0g}+\sigma_{0q} \nonumber\\
&&  \sigma(pp\to\sq_{L,R}\,\sq_{R,L}^*)=\sigma_3/\rap^2 +\mathcal{O}(1/\rap^4) \, , \nonumber\\
&&  \sigma(pp\to\sq_{L,R}\,\sq_{L,R})=\sigma_4/\rap^2 +\mathcal{O}(1/\rap^4) \, , \nonumber\\
&&  \sigma(pp\to\sq_{L,R}\,\sq_{R,L})=\sigma_5/\rap^4 +\mathcal{O}(1/\rap^6) \, ,
\label{forsigi}
\eea
where the coefficients $\sigma_i$ ($i=0g,0q,1\ldots5$) are the same as in \tab{tab:cros}. In \eq{forsigi}, each channel is truncated at its leading order in $1/\rap^2$, with one exception: for the same-chirality $\sq\sq^*$ channel we also retain the $1/\rap^4$ term, since $\sigma_2$ exceeds $\sigma_1$ by more than an order of magnitude (a factor of about 45 at $\msq=1\TeV$), so that the $1/\rap^4$ term dominates for $\rap\lsim5$. Only part of this hierarchy is visible at the partonic level, where the ratio of the two coefficients in \tab{tab:cros} (for $i=j$) is 5 at threshold. The ratio $\sigma_2/\sigma_1$ grows like $\hat s/4\msq^2$ above threshold because the $s$-channel gluon propagator falls as $1/\hat s$, unlike the gluino exchange, and the parton luminosities account for a factor of about 2. Moreover, the interference effect in the $1/\rap^2$ term requires the initial quark--antiquark pair to have the same flavor, whereas the gluino-exchange term receives contributions from all $q_i\bar q_j$ initial states. This accounts for another factor of about 4. 

In the large-$\rap$ regime, the cross-section dependence on the gluino mass is fully described by the powers of $\rap$ shown in \eq{forsigi}, since the cross-section coefficients $\sigma_{i}$ are functions only of $\msq$, and not of $\rap$. In this way, the cross-section dependence on $\msq$ and $\rap$ is completely factorized. This is the reason why we advocate the use of $(\msq ,\rap)$ as input parameters of an improved squark simplified model. Of course, the choice $(\msq ,\mgl)$ is equivalent, but it hides the simple scaling behavior with the dimensionless quantity $\rap$.

\begin{figure}[t!]
\centering
\includegraphics[width=0.9\textwidth]{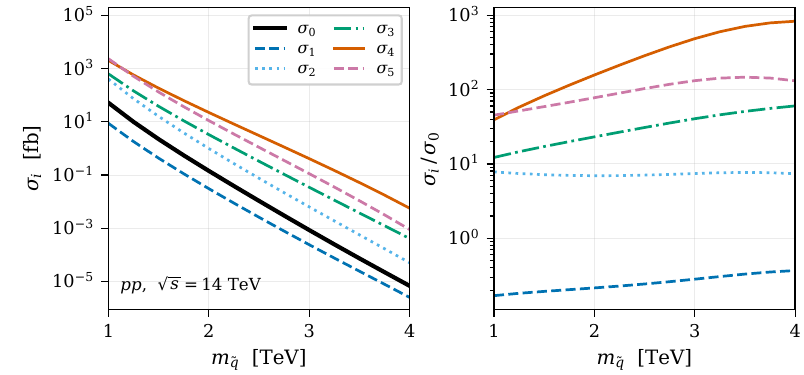}
\caption{The LO cross-section coefficients $\sigma_0\equiv\sigma_{0g}+\sigma_{0q}$ and
$\sigma_{1\ldots5}$, defined in
\eq{forsigi}, as functions of the squark mass, at $\sqrt{s}=14\TeV$ (see text). The coefficient
$\sigma_{4}$ is the largest among all, since $\sq_{L,R}\,\sq_{L,R}$ production is the
only channel that is simultaneously valence--valence and S-wave at threshold. }
\label{fig:scaling}
\end{figure}

The dependence of the coefficients $\sigma_{i}$ on
the squark mass is shown in \fig{fig:scaling}.
The coefficients are obtained by convoluting the partonic cross sections of \tab{tab:cros} with the NNPDF2.3 LO parton distributions~\cite{Ball:2012cx} at $\sqrt{s}=14\TeV$, summing over all initial-state flavors allowed by each process, with renormalization and factorization scales set to $\msq$ and eight first- and second-generation squarks taken to be degenerate. The full cross sections shown in the figures of this section are computed at leading order with \mgfive~\cite{Alwall:2014hca} with the same inputs.

\boldmath
\subsubsection*{Squark--antisquark production: $ \sq \sq^*$}
\unboldmath

In the limit of decoupled gluinos ($\rap \to \infty$), the squark production cross section is given by $\sigma_0=\sigma_{0g}+\sigma_{0q}$, which will be referred to as the ``simplified" cross section, as it corresponds to the reference cross section often used for deriving exclusion limits in collider analyses.

In the processes contributing to the simplified cross section, squarks couple only to gluons and therefore the squark interaction preserves chirality and is flavor universal. This implies that all degenerate squark flavors are produced equally.

The simplified cross section $\sigma_0$ does not fully account for $ \sq \sq^*$ production. Additional contributions come from $t$-channel gluino exchange. A universal property that holds for all the gluino-mediated processes listed in \tab{tab:cros} is that, unlike the simplified contribution, the flavor of the final-state squark matches the flavor of the quark initial state.

Among the different gluino-mediated contributions to $\sq\sq^*$ production, the largest effect comes from $\sigma_3$, since it is the only case with no P-wave suppression and with $1/\rap^2$ asymptotic behavior. The dominance of $\sigma_3$ over $\sigma_{1,2}$ is confirmed by \fig{fig:scaling}.

An important result is that the gluino contribution to  $\sq \sq^*$ production is larger than the simplified contribution as long as
\beq
\rap < \sqrt{\frac{\sigma_3}{\sigma_0}}=
\left\{
\begin{array}{cl}
4.8 & {\textrm{for}}~\msq =2~\tev \\
6.3 & {\textrm{for}}~\msq =3~\tev
\end{array}
\right.
\eeq
This shows that gluino decoupling occurs only at relatively large values of $\rap$
and illustrates the importance of gluinos in squark--antisquark production.

To further quantify the role of gluinos, we show in \tab{tab:decomp} the different contributions to $\sq \sq^*$ production for the particular parameter choice $\msq=3\TeV$ and $\rap =2$. The result is based on a complete calculation and not on the asymptotic behavior, which is unreliable for $\rap =2$. In this specific case, gluinos account for 85\% of the cross section.

\begin{table}[htb]
\centering
\begin{tabular}{lcc}
\toprule
component & $\sigma$ [fb] & fraction of $\sigma( p p \to \tilde{q} \tilde{q}^*)$\\
\midrule
(1) $gg$ fusion & $3.4\times10^{-4}$ & $6\%$\\
(2) $q\bar q$ $s$-channel gluon & $5.1\times10^{-4}$ & $9\%$\\
(3) $q\bar q$ $t$-channel gluino ($+$ interf.) & $5.0\times10^{-3}$ & $85\%$\\
\midrule
simplified (1)+(2), $\mgl\to\infty$ & $8.6\times10^{-4}$ & $15\%$\\
\bottomrule
\end{tabular}
\caption{Contributions to $pp\to\sq\sq^{*}$ from distinct channels. The values correspond to LO cross sections computed at $\sqrt{s} = 14$~TeV, $\msq=3\TeV$ and $\mgl=6\TeV$.}
\label{tab:decomp}
\end{table}

\boldmath
\subsubsection*{Di-squark production: $ \sq \sq$}
\unboldmath

Squark--squark production occurs only through gluino exchange and therefore it is not taken into account by the squark simplified model. Nonetheless, the $\sq\sq$ channel gives a very large contribution to squark production, as shown in \fig{fig:scaling}. The reason why di-squark production is so large is that it is initiated by two valence quarks, while $\sq\sq^{*}$ needs valence--sea quarks or gluon--gluon initial states.

Indeed, \fig{fig:scaling} shows that $\sigma_4$ is the largest among all cross-section coefficients. This can be easily understood from \tab{tab:cros} by noting that the same-charge/same-chirality channel is the only one benefitting from {\it (i)} valence--valence quark PDF and {\it (ii)} S-wave production at threshold.

The suppression of $\sigma_5$ with respect to $\sigma_4$ can be attributed to the different nature of the leading partial wave.
Numerically the penalty is mild and, at the lowest squark masses, it is not a penalty at all. Measured at fixed $\shat$ with the gluino far off shell, so that the propagator carries no $\shat$ dependence of its own, the two threshold behaviors are:
 $\hat\sigma\propto\beta$ in S-wave and $\hat\sigma\propto\beta^{3}\gamma^{2}$ in P-wave, with $\gamma^{2}=\shat/4\msq^{2}$. Their ratio is $p^{*2}/\mgl^{2}$, where $p^{*}$ is the momentum of each squark in the partonic center-of-mass frame, $p^{*2}=\shat\beta^{2}/4=\msq^{2}\beta^{2}\gamma^{2}$. The $\mgl^{2}$ cancels the $\rap^{2}$ that the coefficients in \eq{forsigi} already carry, so the kinematic P/S factor is $\langle p^{*2}\rangle/\msq^{2}$, which is equal to $1.59$, $0.63$, $0.33$, $0.19$ for $\msq=1,2,3,4\TeV$. This can be compared with the results in \fig{fig:scaling}, which yield
 $\sigma_{5}/\sigma_{4}=1.16$, $0.50$, $0.28$, $0.16$ for the same values of the squark mass. The discrepancy between the two determinations is explained by channel multiplicity and color factors, rather than kinematics. So, the P-wave channels are reduced by a factor of a few at high mass, but the suppression \emph{inverts} below $\msq\simeq1.2\TeV$, where the collisions are energetic enough that $\langle p^{*2}\rangle>\msq^{2}$, and the non-relativistic expansion fails. This explains why $\sigma_{5}$ is larger than $\sigma_{4}$ for $\msq =1\TeV$, see \fig{fig:scaling}.

A measure of the importance of gluinos in di-squark production is that its contribution to the cross section remains larger than the simplified value as long as
\beq
\rap < \sqrt{\frac{\sigma_4}{\sigma_0}}=
\left\{
\begin{array}{cl}
13 & {\textrm{for}}~\msq =2~\tev \\
22 & {\textrm{for}}~\msq =3~\tev
\end{array}
\right.
\eeq
The effect is remarkable: even gluinos 10 or 20 times heavier than quarks give the dominant contribution to squark production.

\begin{figure}[htb]
\centering
\includegraphics[width=0.86\textwidth]{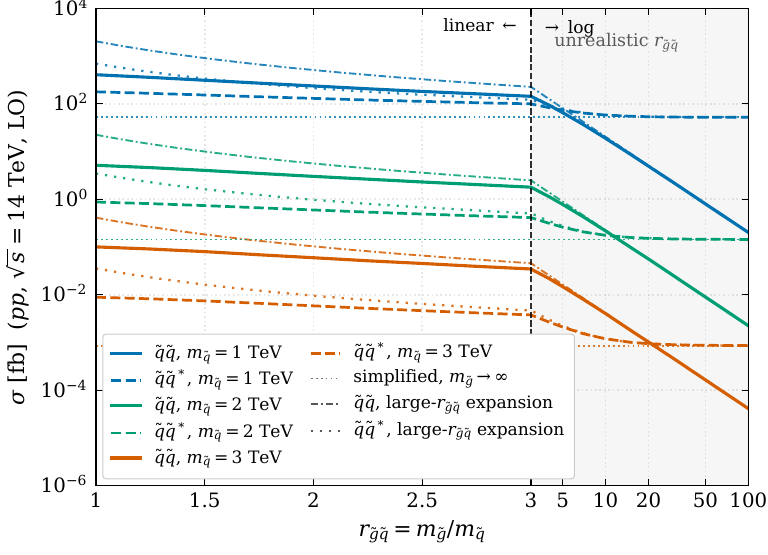}
\caption{LO squark production cross sections at $\sqrt{s} = 14$~TeV versus $\rap=\mgl/\msq$ for $\msq=1,2,3\TeV$ (different colored lines). Solid:
 $\sq\sq$; dashed: $\sq\sq^{*}$. Dotted color-matched lines correspond to the
$\mgl\to\infty$ simplified limit of $\sq\sq^{*}$ production. The axis is linear
below $\rap=3$ and logarithmic above (dashed divider). The shaded region corresponds to $\rap>3$. Thin color-matched lines show the leading terms in the large-$\rap$ expansion of \eq{forsigi} for $\sq\sq$ (dash-dotted) and $\sq\sq^*$ (loosely dotted).
}
\label{fig:asy}
\end{figure}

\boldmath
\subsection{The realistic range $1<\rap\lsim 3$ \label{sec:realistic-r-range}}
\unboldmath

The study of the large-$\rap$ behavior was useful in inspecting the structure of the squark production cross section, but it refers to a parameter region of no interest for low-energy supersymmetry. As discussed in \sec{sec:ratio}, realistic supersymmetric theories extend only up to about $\rap\approx 2$--3. This region is the focus of this section.

For $1<\rap\lsim 3$, the categorization of the squark production channels in \tab{tab:cros} is still valid, but 
the expansion in $1/\rap$ is unreliable.

The complete result to all orders in $1/\rap$ is shown in \fig{fig:asy}. 
The dotted lines show the simplified model cross sections ($\sigma_0$)  for $\msq=1,2$ and 3~TeV. A comparison with the corresponding fine dotted lines shows the relevance of the gluino contribution to the production cross section, which is dominant for all realistic values of $\rap$. Figure~\ref{fig:asy} also shows the large $\rap$ expansion as dot-dashed and dotted lines. We can see that these curves reproduce well the behavior of the full cross sections for sufficiently high $\rap$. In particular, for $\msq=3$~TeV, the expansion is a good approximation for $\rap \gtrsim 3$. For $\rap \lesssim 3$, the large-$\rap$ expansion significantly overestimates the cross sections.
In this regime, which is the relevant one for realistic supersymmetric theories, the full expression for the cross sections (solid and dashed lines) must be used.

\begin{figure}[t!]
\centering
\includegraphics[width=\textwidth]{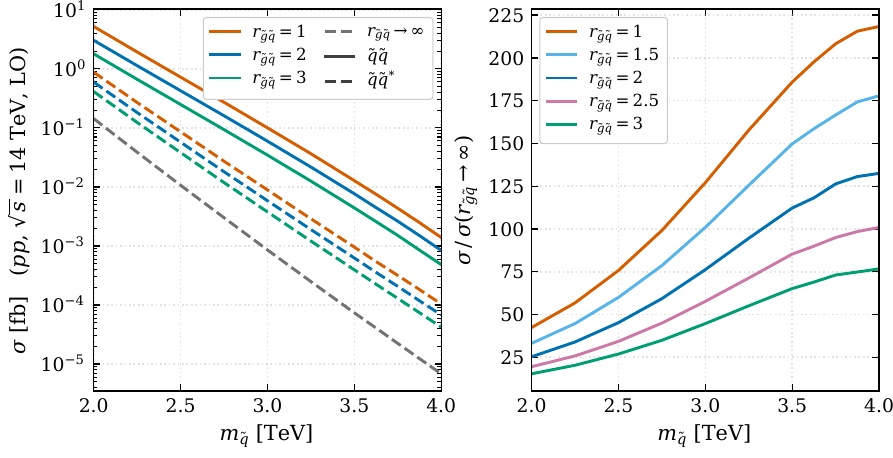}
\caption{Left: Squark production cross sections at $14\TeV$ (LO), at fixed $\rap$, as a function of $\msq$. Solid: $\sq\sq$. Dashed: $\sq\sq^{*}$. Different colored lines refer to different values of $\rap$. The simplified-model result ($\rap \to \infty$ in gray) is entirely due to the $\sq \sq^*$ channel. Right: Ratio of the total squark production cross section $\sigma=\sigma(\sq\sq)+\sigma(\sq\sq^*)$ at finite $\rap$ over the simplified-model prediction ($\rap\to\infty$), for $\rap=1$, 1.5, 2, 2.5 and 3.}
\label{fig:main}
\end{figure}

The left panel of \fig{fig:main} shows the dependence of the squark production cross section as a function of $\msq$, for realistic values of $\rap$. The solid (dashed) lines correspond to the $\sq \sq$ ($\sq \sq^*$) channels. As expected, the di-squark cross section is well above squark--antisquark for $\rap < 3$ and $\msq \geq 1$~TeV. Furthermore, the gap between the $\tilde{q} \tilde{q}$ and $\tilde{q} \tilde{q}^*$ cross sections increases with $\msq$, due to
parton luminosity. The ratio ${\cal L}_{qq}/({\cal L}_{q\bar q}+{\cal L}_{gg})$, evaluated at
$\shat=4\msq^{2}$, grows from $5$ at $\msq =2$~TeV to $18$ at $\msq =4$~TeV. The ratio of cross sections tracks these values, with deviations of less than 30\%, which can be attributed to differences in the matrix elements. The enhancement of the squark production cross section with respect to the simplified model is illustrated in the right panel of \fig{fig:main}.

\begin{figure}[htb]
\centering
\includegraphics[width=\textwidth]{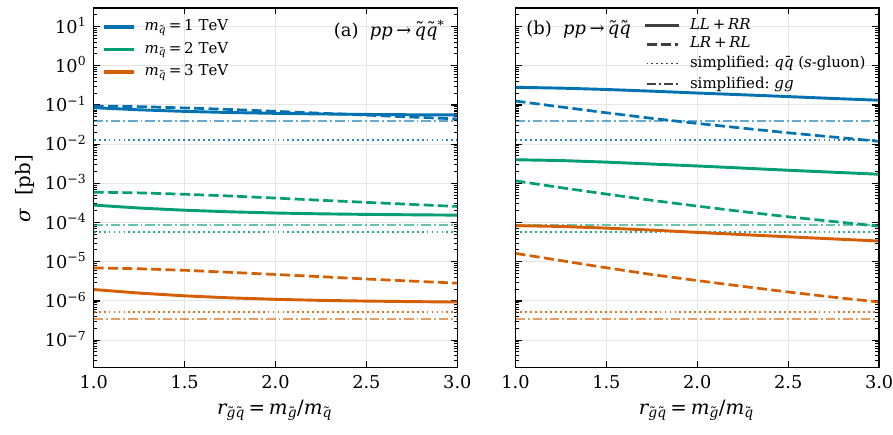}
\caption{Chirality decomposition of the LHC squark production cross section versus $\rap=\mgl/\msq$, summed over the flavors of the first two generations, for $\msq=1,2,3\TeV$ (colors). Left (a): $pp\to\sq\sq^{*}$, where $LL+RR$ stands for $\sq_L\sq_L^*+\sq_R\sq_R^*$ and $LR+RL$ for $\sq_L\sq_R^*+\sq_R\sq_L^*$; right (b): $pp\to\sq\sq$. Solid: $LL+RR$. Dashed: $LR+RL$. Dotted and dash-dotted: the $q\bar q$ $s$-channel-gluon and $gg$-fusion parts of the simplified-model rate. In (a), $LR+RL$ is pure gluino exchange falling as $1/\mgl^2$, while $LL+RR$ flattens onto the simplified value. In (b), $LL+RR$ falls as $1/\mgl^2$ (Majorana insertion) and $LR+RL$ as $1/\mgl^4$.
}
\label{fig:chir}
\end{figure}

The decomposition of the squark production cross section into the different chiral channels is shown in \fig{fig:chir}. 
The behavior of each chirality component confirms the expectation from the expressions in  \tab{tab:cros}. 
First, we notice that the P-wave components, which correspond to the $LR+RL$ (dashed) curves in $p p \to \sq \sq^*$ and to the $LL+RR$ (solid) curves in $p p \to \sq \sq$, are always sub-dominant with respect to the corresponding S-wave contributions. This is due to the $\beta^3$ (P-wave) suppression at large $\msq$.
The only exception is the $\msq = 1$~TeV
$\sq \sq^*$ curve, 
where the P-wave suppression is less relevant and the same-chirality contribution becomes dominant for $\rap \gtrsim 2.5$.
Second, as expected from the previous discussion, the $\sq \sq$ channel, which requires a gluino $t$-channel exchange, is significantly larger than the $\sq \sq^*$ process, for all realistic values of $\rap$.

In conclusion, gluinos play a key role in the squark production cross section. Their contributions are so much larger than the simplified cross section that they continue to give the dominant contribution at the LHC, even if their mass were 10--20 times larger than $\msq$. 
When we combine this result with the theoretical observation that realistic supersymmetric models predict that $\rap$ cannot exceed about 2--3, we reach the conclusion that the squark simplified model completely misrepresents the squark production cross section. 

On several accounts, the squark simplified model fails to describe the production cross section. First, it omits the di-squark channel, which gives by far the dominant contribution, since it originates from valence quarks. Second, it includes the squark--antisquark channel, but misses its leading contribution.

The discrepancy between the realistic value of the cross section and the value predicted by the simplified model grows with $\msq$. Therefore, including the gluino contribution becomes increasingly important as searches accumulate integrated luminosity, which will be the case with HL-LHC.

\section{Implications for squark searches at colliders}
\label{sec:reach}

In this section, we study the impact of the gluino--squark mass ratio $\rap$ on squark searches at current and future colliders.

\subsection{LHC}

The total squark production cross section, studied in \sec{sec:anatomy},
directly impacts the limits on the squark mass.
To translate the values of the cross section into limits on $\msq$, we re-interpret the 13~TeV $\mathcal{L}=$139~fb$^{-1}$  ATLAS $0$-lepton${}+{}$monojet squark search reported in ref.~\cite{ATLAS:2020syg} and extrapolate it to the case of finite $\rap$. The main result we use is the ATLAS
95\% C.L. upper limit on the production cross section ($\sigma_{UL}$) as a function of $\msq$ and $\mlsp$ for the squark simplified model
(the  recasting material is described in detail in \app{app:materials}). 

Adding the parameter $\rap$ significantly changes the production rate, but does not modify the decay mode, since $\sq\to q\gl$ is kinematically closed for $\rap >1$. Moreover, for $\msq$ close to the exclusion limit, $\rap$ has a limited impact on kinematic distributions and experimental acceptance (more details in \ref{app:caveats}). Therefore, we can produce mass exclusion plots by requiring 
\begin{equation}
\sigma_{\textrm{tot}}(\msq,\,\rap)\;<\;\sigma_{UL}(\msq,\,\mlsp)\, .
\label{eq:recast}
\end{equation}
Here $\sigma_{\textrm{tot}}=\sigma_{\sq\sq}+\sigma_{\sq\sq^{*}}$ 
is the theoretical prediction for the total cross section discussed in \sec{sec:anatomy}. For finite $\rap$, the cross section $\sigma_{\textrm{tot}}$ is larger than the simplified model value ($\sigma_0$), leading to stronger bounds on the squark mass.

We compute $\sigma_{\textrm{tot}}$ at NNLO$_{\textrm{approx}}$+NNLL\cite{Beenakker:2016lwe,Beenakker:2024jwh} with the
PDF4LHC15 \cite{Butterworth:2015oua} set used by ATLAS (details of our calculation and necessary extrapolations are given in \app{app:cross-section}).
This allows us to derive, for $\rap\to \infty$, the squark mass limit shown in the left panel of \fig{fig:reach}
(gray dotted line), which reproduces quite well the original ATLAS result (black solid line).

When we analyze the case $1<\rap\lsim 3$, the exclusion moves to squark masses above
$2.4\TeV$, which is the maximum value of $\msq$ for which ATLAS has reported data on $\sigma_{UL}$. Therefore, we must extrapolate our analysis into a mass region where we do not have direct information on $\sigma_{UL}$. 
To overcome this limitation, we use 
{points taken from Fig.~17 of 
ref.~\cite{ATLAS:2020syg} at the values of $\rap$ under study.}

Extrapolating from these points, we obtain the exclusion lines corresponding to $\rap =1,2,3$, which are presented in the left panel of \fig{fig:reach}. Needless to say, our results can only be regarded as an estimate of the mass limits. While we believe that our estimates are fairly reliable, a complete data analysis can only be performed by the experimental collaborations.

\subsubsection*{The $\rap=1$ edge}
As long as $\rap$ is sufficiently large, we avoid contamination from direct $\gl\gl$ gluino-pair production, $\sq\gl$ associated production, or decay $\sq\to q\gl$. Indeed, for $\rap \ge 2$, the $\gl\gl$ and $\gl\sq$ processes amount to at most $10^{-3}$ times the squark rate, and are therefore negligible. However, for $\rap \approx 1$, the gluino production channels cannot be ignored. Therefore, the case $\rap \approx 1$ requires a special treatment.

\begin{figure}[t!]
\centering
\includegraphics[width=\textwidth]{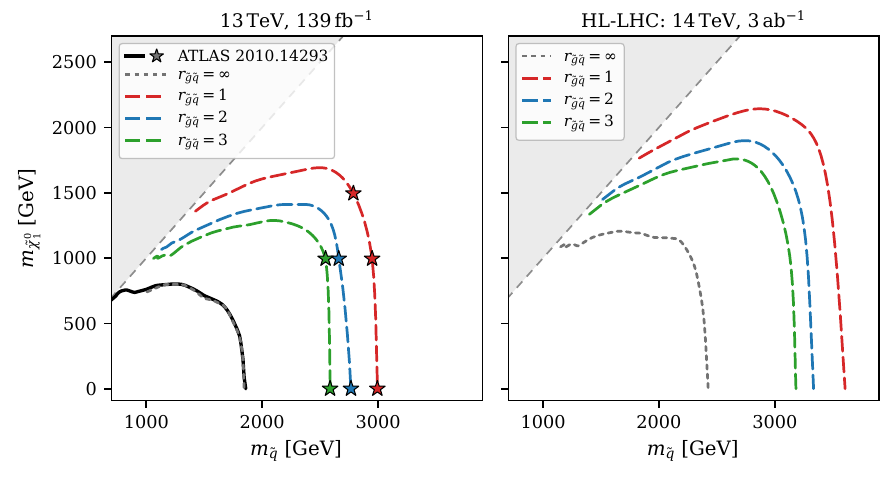}
\caption{Exclusion contours in the $\msq$--$\mlsp$ plane obtained with the procedure described in the text. The left panel refers to current LHC data and the right panel to HL-LHC projections. For the LHC, we also show the ATLAS limit for the squark simplified model (solid line from \cite{ATLAS:2020syg,https://doi.org/10.17182/hepdata.95664.v2/t46}) and the points with finite gluino mass that have been excluded by ATLAS (marked with stars). The compressed region (very small $\Delta=\msq-\mlsp$) is {affected by larger uncertainties from extrapolation}  (see \app{app:caveats}).}
\label{fig:reach}
\end{figure}

In order to compute an exclusion curve for $\rap=1$, we assume the same acceptance and efficiency for the $\sq\sq^*$, $\sq\sq$, $\sq\gl$ and $\gl\gl$ channels. For $\sq\gl$ and $\gl\gl$, this is  clearly an approximation, since the gluino decay products have distinct kinematics from the squark ones.
The result is shown by the red curve in the left panel of \fig{fig:reach}. Since our extrapolation of $\sigma_{UL}$ is anchored to the exclusions measured by ATLAS at $\rap=1$ (red stars), which carry the correct gluino efficiencies, the approximation is largely absorbed near these points, and its residual effect on the mass limit is at most a few tens of GeV.
The relative contributions from each production mode, for $\msq=3\TeV$ (which corresponds to the mass limit for $\mlsp=0$ and $\rap = 1$), are $\sq\sq:65\%$, $\sq\gl:29\%$, $\sq\sq^{*}:4\%$ and
$\gl\gl:2\%$.
Note that gluino exchange dominates also the $\sq\sq^*$ channel. With the gluino decoupled, as in the simplified model, this channel would amount to only $0.5\%$ of the total production rate at $\rap=1$.

\begin{table}[t!]
\centering
\setlength{\tabcolsep}{5pt}
\begin{tabular}{lcc| cc| cc| cc}
\toprule
 & \multicolumn{2}{c}{LHC} & \multicolumn{2}{c}{HL-LHC} & \multicolumn{2}{c}{FCC-hh} & \multicolumn{2}{c}{FCC-hh}\\
 & \multicolumn{2}{c}{$13\TeV$, $139\fb^{-1}$} & \multicolumn{2}{c}{$14\TeV$, $3\ab^{-1}$} & \multicolumn{2}{c}{$85\TeV$, $30\ab^{-1}$} & \multicolumn{2}{c}{$100\TeV$, $30\ab^{-1}$}\\
\cmidrule(lr){2-3}\cmidrule(lr){4-5}\cmidrule(lr){6-7}\cmidrule(lr){8-9}
\\[-8mm]
 & $\msq$ limit & & $\msq$ reach & & $\msq$ reach & & $\msq$ reach &
 \\[-6mm]
$\rap$ &  & {\large $\frac{\sigma_{\textrm{tot}}}{\sigma_{0}}$} &  & {\large $\frac{\sigma_{\textrm{tot}}}{\sigma_{0}}$} &  & {\large $\frac{\sigma_{\textrm{tot}}}{\sigma_{0}}$} &  & {\large $\frac{\sigma_{\textrm{tot}}}{\sigma_{0}}$}
\\[-4.4mm]
 & [TeV] & & [TeV] & & [TeV] & & [TeV] & 
 \\[0mm]
\midrule
$\infty$ & $1.8$ & $1$ & $2.4$ & $1$ & $12$ & $1$ & $14$ & $1$\\
$3$  & $2.6$ & $26$ & $3.2$ & $33$ & $18$ & $46$ & $20$ & $43$\\
$2$  & $2.8$ & $55$ & $3.3$ & $70$ & $19$ & $89$ & $22$ & $85$\\
$1$  & $3.0$ & $195$ & $3.6$ & $214$ & $20$ & $215$ & $23$ & $207$\\
\bottomrule
\end{tabular}
\caption{The current LHC limit on the squark mass and the reach at future colliders, in the case of massless $\lsp$. Also shown is $\sigma_{\textrm{tot}}/\sigma_{0}$ (computed at the corresponding values of $\msq$ and $\rap$), which indicates the enhancement factor with respect to the simplified model. The results include higher-order QCD corrections (\nnllfast, see Appendix~\ref{app:cross-section}), which are larger for $\sq\sq^*$ than for $\sq\sq$ and therefore reduce $\sigma_{\textrm{tot}}/\sigma_{0}$ by about 40\% with respect to the LO ratios shown in the right panel of \fig{fig:main}.}
\label{tab:reach}
\end{table}
\subsubsection*{Results}
As shown in \fig{fig:reach} and \tab{tab:reach}, a finite gluino mass has a significant impact on the squark mass limit. The gain in squark mass with respect to the simplified model is $1\TeV$ for
$\rap=2$ and $1.2\TeV$ for $\rap=1$. These mass shifts are far larger than the uncertainties associated with our extrapolation procedure to explore the squark mass region in which ATLAS has not provided direct information about $\sigma_{UL}$. Therefore, despite the approximations we make, our conclusions are robust.

\subsection{HL-LHC}
\label{sec:hl}
We can extend our analysis and make projections for HL-LHC, based on $\sqrt{s}=14\TeV$ and a final integrated luminosity of 3~ab$^{-1}$. We employ the same procedure used for the LHC, while accounting for the luminosity increase and change of the background rate with energy (see \app{app:projections} for more details). Our results are shown in the right panel of \fig{fig:reach} and in \tab{tab:reach}.

Our projection predicts a squark-mass reach at HL-LHC of 2.4~TeV for the simplified model ($\rap\to \infty$), which is in good agreement with ref.~\cite{deBlas:2025gyz}. When we turn to realistic values of $\rap$, we find that the simplified model underestimates the squark production rate by a factor of 70 (for $\rap=2$) and more than 200 (for $\rap=1$). The enhancement factors at HL-LHC are larger than those at LHC because the increase in luminosity allows us to probe larger $\msq$.

The gain in squark-mass reach with respect to the simplified model is $0.9\TeV$ at $\rap=2$ and $1.2\TeV$ at $\rap=1$. This mass gain is essentially identical to what we found for the LHC. This is because the cross section steeply falls with $\msq$, while the experimental upper limit $\sigma_{UL}$ improves only slowly with luminosity. The mass reach grows only logarithmically with the enhancement factor and, therefore, its increase at HL-LHC corresponds to a rather small shift in the mass gain.

Our projections are based on a simplified approach, whose details and limitations are discussed in \app{app:projections} and \app{app:caveats}. Definitive results require a full treatment that can only be performed by the experimental collaborations.

\subsection{FCC-hh}
\label{sec:fcc}
Next, we repeat our analysis for future high-energy colliders, taking FCC-hh as the prototype. For the $pp$ center-of-mass energy, we consider both the baseline value of 85 TeV and the long-term goal of 100 TeV. We derive limits by rescaling the luminosity \emph{and} mass scales. 
 For a search whose selection criteria track
the signal mass, we can assume that the surviving background and signal acceptance depend on
$\msq$ and $\sqrt{s}$ only through the combination $\tau=4\msq^2/s$. In this way, we can translate the ATLAS squark-mass limit into the FCC-hh reach on a squark mass that corresponds to the same value of $\tau$ (see \app{app:projections} for more details on the rescaling procedure and \app{app:caveats} for possible limitations of our approach). 

\begin{figure}[t!]
\centering
\includegraphics[width=\textwidth]{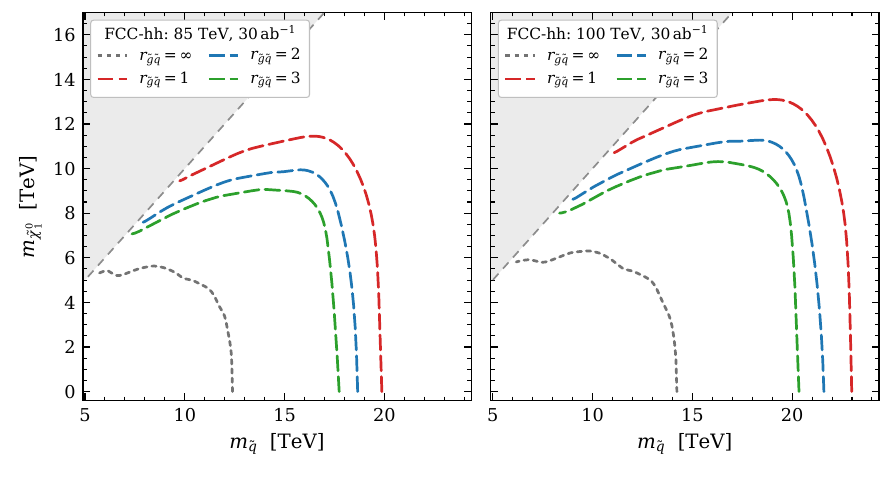}
\caption{Exclusion contours in the $\msq$--$\mlsp$ plane for FCC-hh with $30\ab^{-1}$, and $\sqrt{s}=85\TeV$ (left panel) or $\sqrt{s}=100\TeV$ (right panel). The vertical axis is identical in both panels.
 The compressed region (very small $\Delta=\msq-\mlsp$) {is affected by larger uncertainties from extrapolation} (see \app{app:caveats}). }
\label{fig:fccreach}
\end{figure}

Our results are shown in \fig{fig:fccreach} and \tab{tab:reach}.
We estimate that the simplified model predicts a squark-mass reach at the FCC-hh of 12 TeV, in agreement within 4\% with ref.~\cite{deBlas:2025gyz}.
For realistic values of $\rap$, the squark production cross section is enhanced with respect to the simplified model by a factor of 90 at $\rap =2$, and more than 200 at $\rap =1$. Therefore, the enhancement factors are essentially the same as for HL-LHC. The mass reach is significantly improved  with respect to the simplified model. We find that FCC-hh is sensitive to squark masses up to 19 TeV for $\rap=2$ (with a mass gain of 7 TeV) or 20 TeV for $\rap=1$ (with a mass gain of 8 TeV). This shows that the simplified model grossly underestimates the physics potential of FCC-hh in the search for squarks.

\section{Impact of electroweak gauginos \label{sec:ew-inos}}

The slow decoupling of the gluino contribution to the squark production cross section is primarily a consequence of the Majorana nature of the gluino. A similar effect is therefore also present in the contributions from $t$-channel diagrams involving EW gauginos (Wino and Bino). There are, however, some important differences with respect to the gluino effect.
\begin{itemize}
\item{EW gaugino contributions are suppressed, with respect to the gluino contribution, by the ratio of EW over QCD couplings.}
\item{Unlike gluinos, EW gauginos have chiral-dependent couplings. In particular, Winos couple only to left-squarks. However, as same-chirality $qq\to \sq_{L,R}\, \sq_{L,R}$ is the leading  channel, the dominant contribution to squark production is present also for Winos, although with only half of the possible channels.}
\item{As the LHC mass bounds on EW gauginos are weaker than those on gluinos, Winos can be lighter than squarks. This can possibly enhance their virtual effect. The decay channel $\sq \to q^\prime \tilde{W}^\pm$ can also become kinematically accessible.}
\item{Typically, the Bino is assumed to be lighter than squarks to allow for the decay $\sq \to q \lsp$. For consistency, its virtual contribution to the squark production cross section must be included.}
\end{itemize}

\begin{figure}[t!]
\centering
\includegraphics[width=0.48\textwidth]{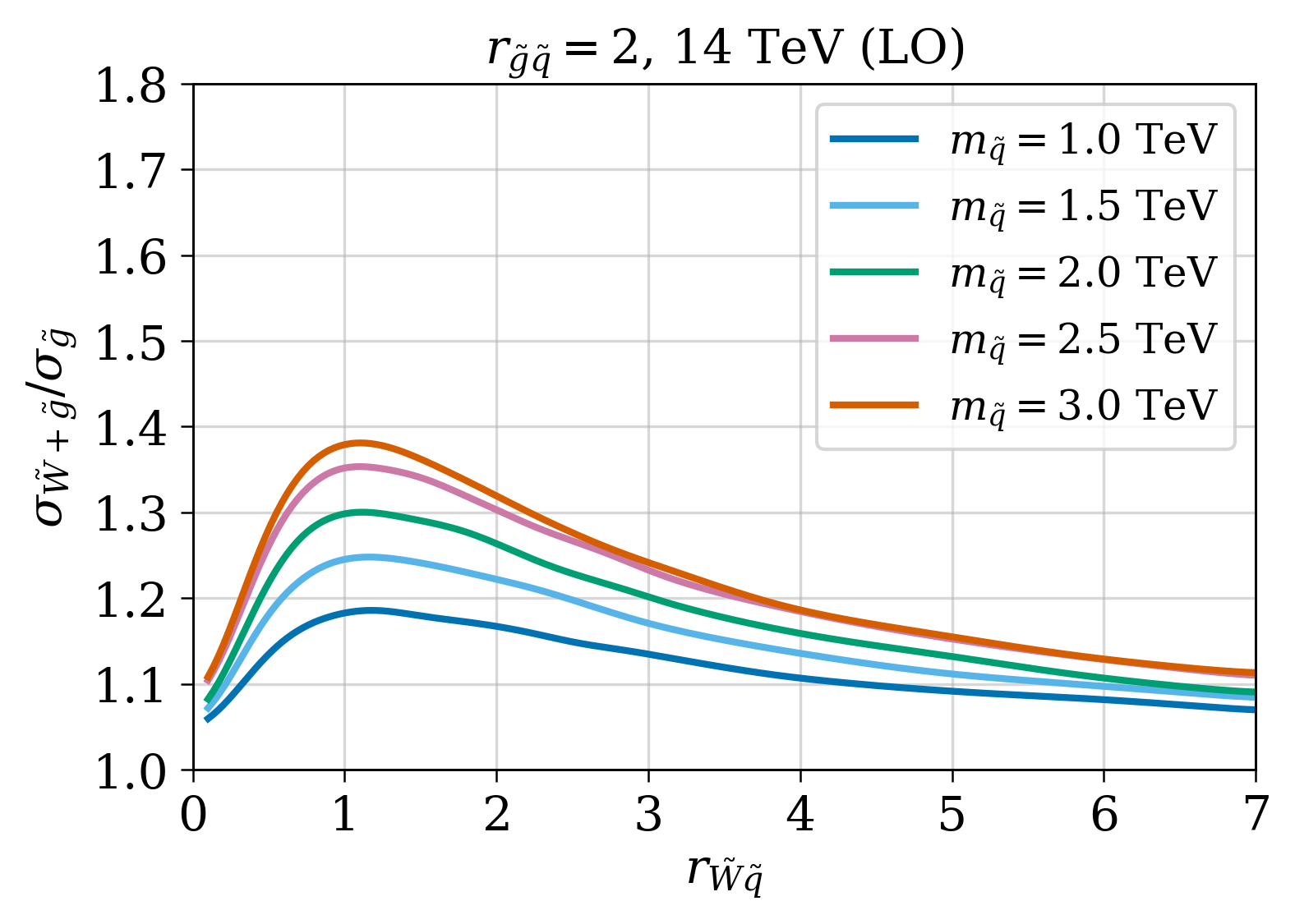}
\includegraphics[width=0.48\textwidth]{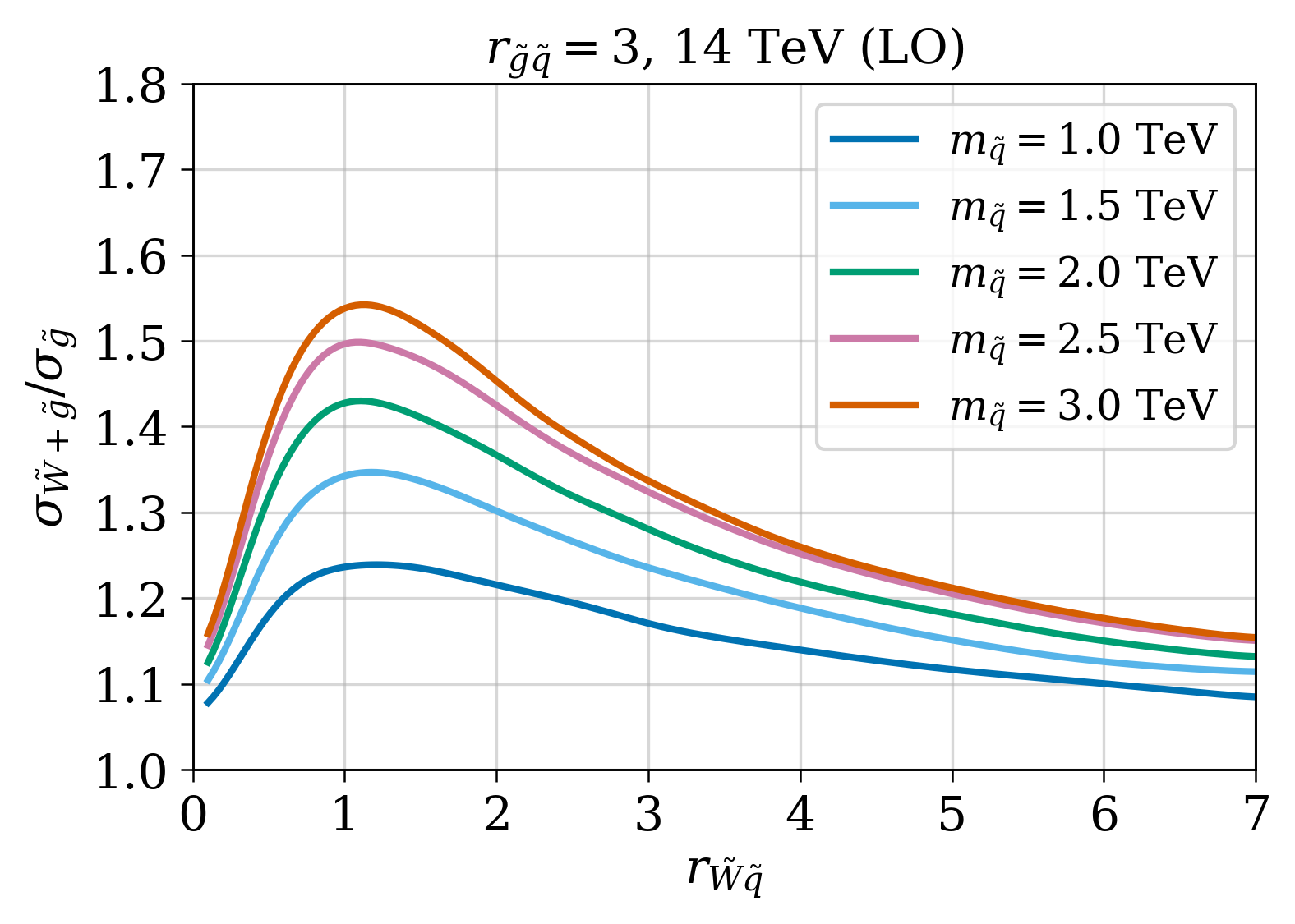}
\includegraphics[width=0.48\textwidth]{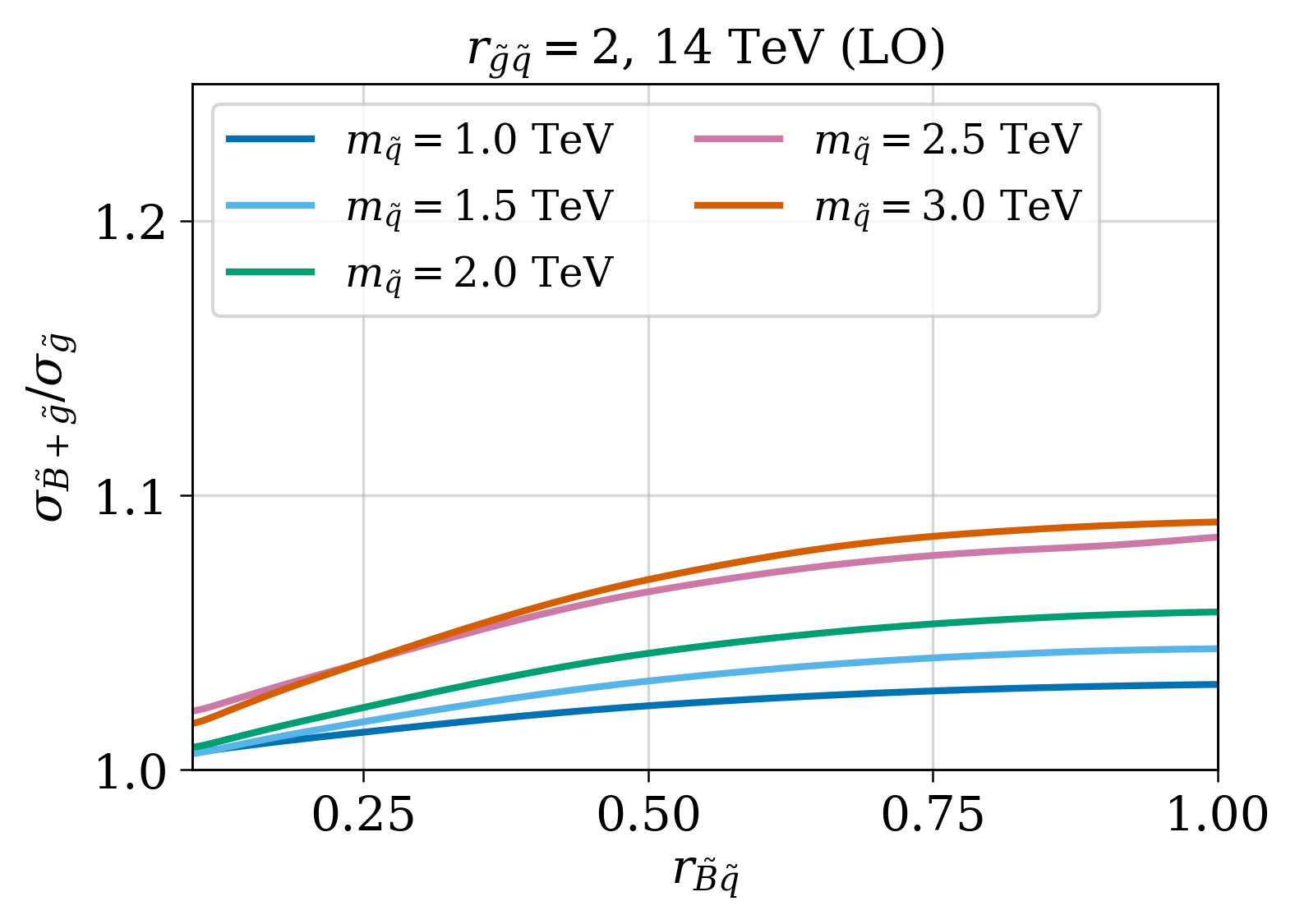}
\includegraphics[width=0.48\textwidth]{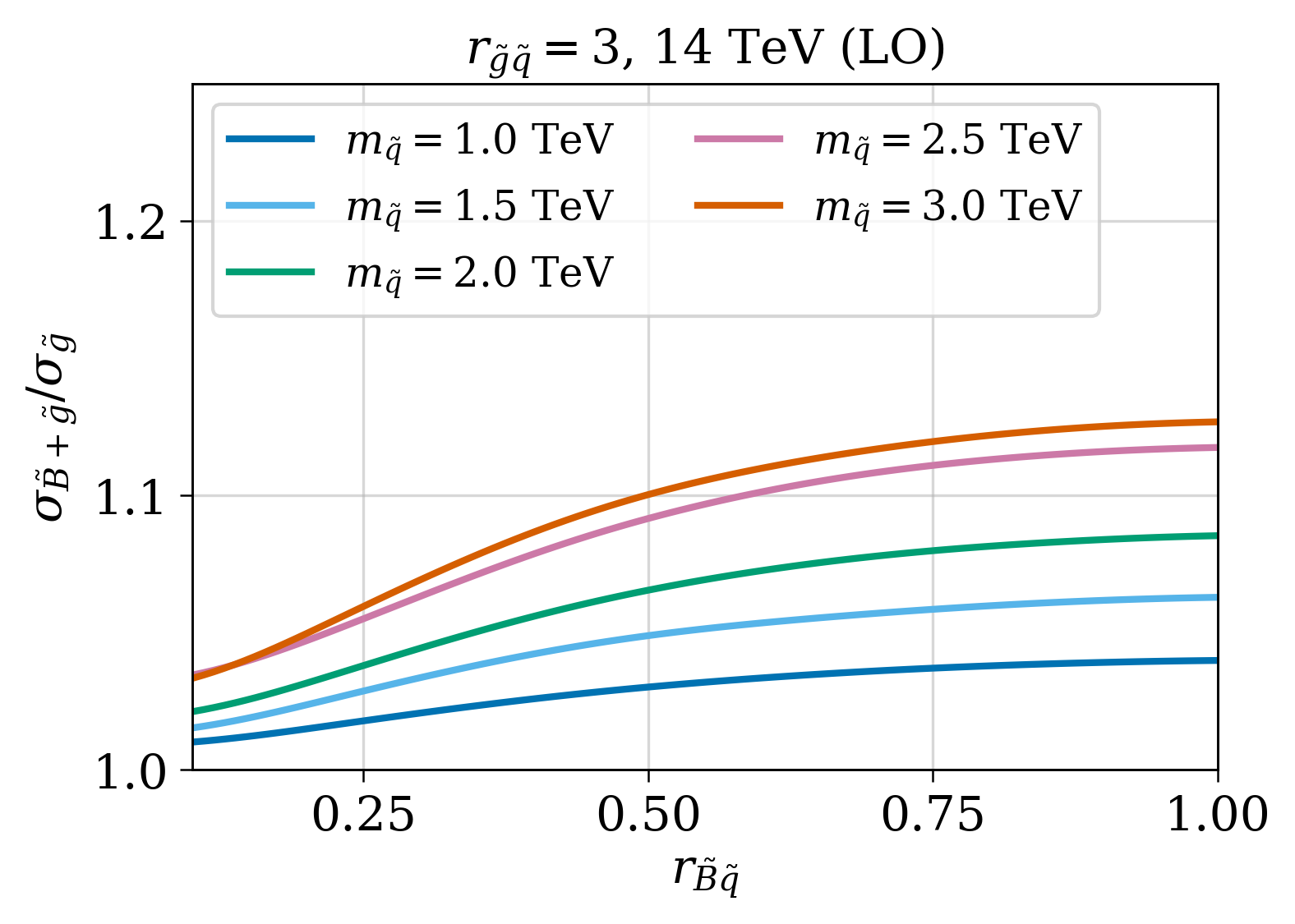}
\caption{LO squark production cross sections at 14 TeV including both the Wino and gluino contributions ($\sigma_{\tilde{W}+\tilde{g}}$, top panels) or Bino and gluino contributions ($\sigma_{\tilde{B}+\tilde{g}}$, bottom panels), in units of the cross section including only the gluino contribution ($\sigma_{\tilde{g}}$). These ratios are shown
as functions of $\rapW= m_{\tilde{W}}/m_{\tilde{q}}$ (top panels) or $\rapB= m_{\tilde{B}}/m_{\tilde{q}}$ (bottom panels) for several values of $m_{\tilde{q}}$, and for $\rap=2$ (left panels) and $\rap=3$ (right panels).}
\label{fig:xsecRatioWino}
\end{figure}

In the mass range relevant for our considerations, EW gauginos consist of nearly pure Winos ($m_{\tilde\chi_2^0} = m_{\tilde\chi_1^\pm}\, = m_{\tilde{W}}$) and Bino ($\mlsp = m_{\tilde{B}}$) states.
Following the same notation introduced above, we define
\beq
    r_{\chi \tilde{q}} \equiv \frac{m_{\chi}}{m_{\tilde{q}}} \, , ~~~~~ \chi = \tilde{B},\,\tilde{W} \, .
\eeq

Let us first consider the Wino contribution to the main squark production channel $qq\to \sq_{L}\, \sq_{L}$. When $\rapW\ll1$, the matrix element is suppressed by the chiral flip and goes like $\alpha_W \rapW$. When $\rapW\gg1$, the matrix element is suppressed by the mass in the propagator and goes like $\alpha_W /\rapW$. This shows that the Wino contribution has a maximum around $\rapW\approx 1$. Moreover, interference with gluinos gives the leading $\rapW$-dependent contribution, with the exception of the region $\alpha_s /\alpha_W\lsim \rap \lsim 3$ and $\rapW \approx 1$. Therefore, the additional Wino contribution to the squark production rate, in units of the pure gluino contribution, has the following asymptotic parametric behavior
\beq
    \frac{\sigma_{\tilde{W}}}{\sigma_{\tilde{g}}} \sim 
    \left\{
    \begin{array}{cl}
    (\alpha_W/{\alpha_s})\, (\rap \, \rapW)  & {\textrm{for}}~\rapW \ll 1 \\[2mm]
     (\alpha_W/{\alpha_s})\, (\rap / \rapW) & {\textrm{for}}~\rapW \gg 1
    \end{array}
    \right.  
\eeq
Here $\alpha_W/{\alpha_s} \approx 0.4$ at the squark mass scale, $\sigma_{\tilde{W}}$ corresponds to the Wino contribution including the interference terms and $\sigma_{\tilde{g}}$ to the pure gluino cross section. 

The top panels of \fig{fig:xsecRatioWino} show the ratio of the total cross section including both the Wino and gluino contributions over the pure gluino cross section for $\rap =2$ and 3. The enhancement factor increases with $m_{\tilde{q}}$, as expected from the discussion in \sec{sec:anatomy}. In realistic supersymmetric scenarios with $\rap \lsim 3$, the Wino contribution can be at most 50\% of the gluino contribution. Therefore, the Wino effect has a subdominant impact on the production cross section and a negligible impact on the squark mass bound.

The case of the Bino is completely analogous, but the impact on the squark production cross section is even smaller, resulting in an enhancement around $10$\% or below, as shown in the bottom panels of \fig{fig:xsecRatioWino}.

\subsubsection*{Impact of electroweak gauginos on the squark simplified model}

Although the squark simplified model is not adequate to describe low-energy supersymmetry, it is instructive to consider how its predictions are affected by EW gauginos. This is particularly relevant for the Bino, since its mass is one of the parameters of the simplified model.

\begin{figure}[t!]
\centering
\includegraphics[width=0.48\textwidth]{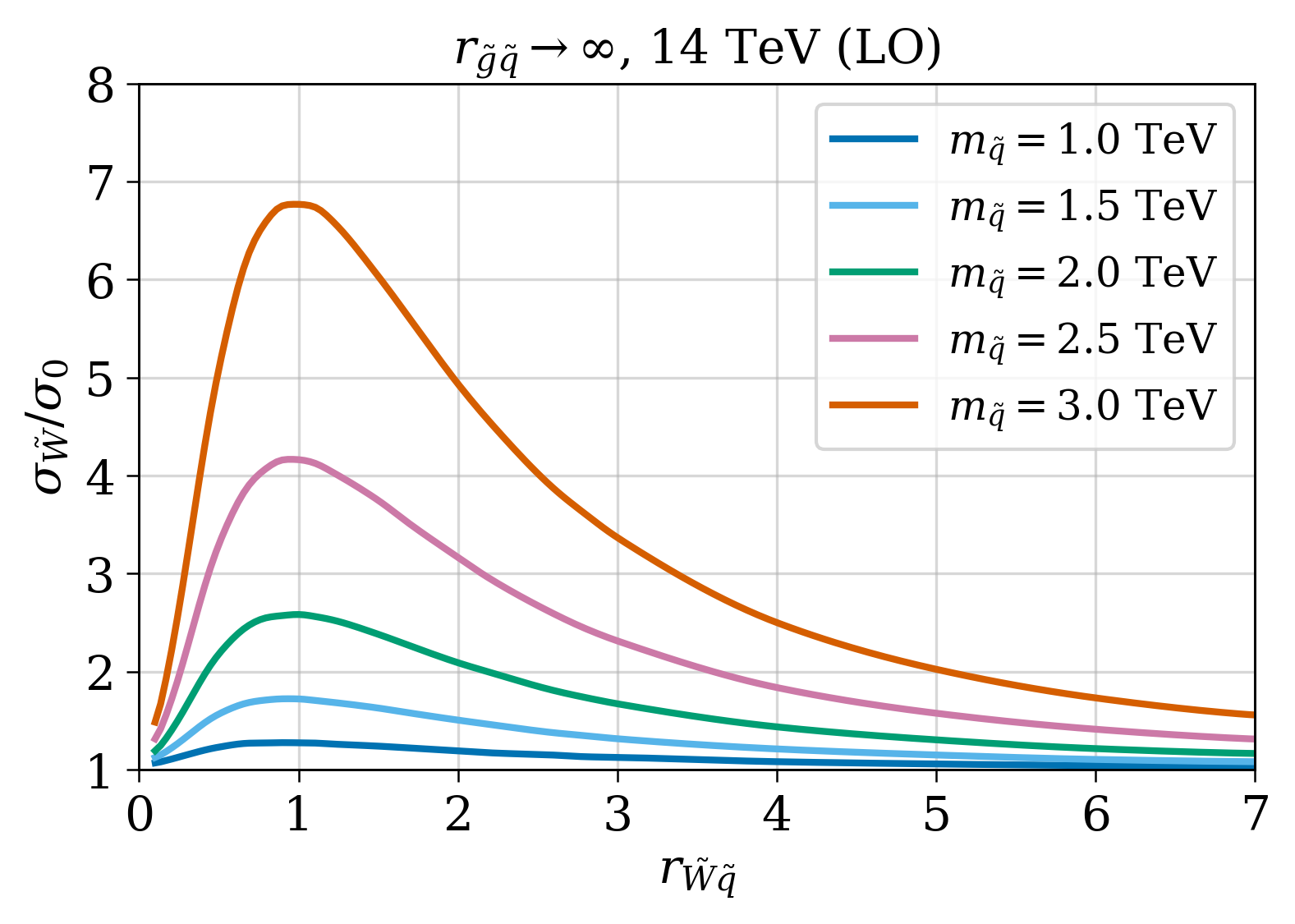}
\includegraphics[width=0.48\textwidth]{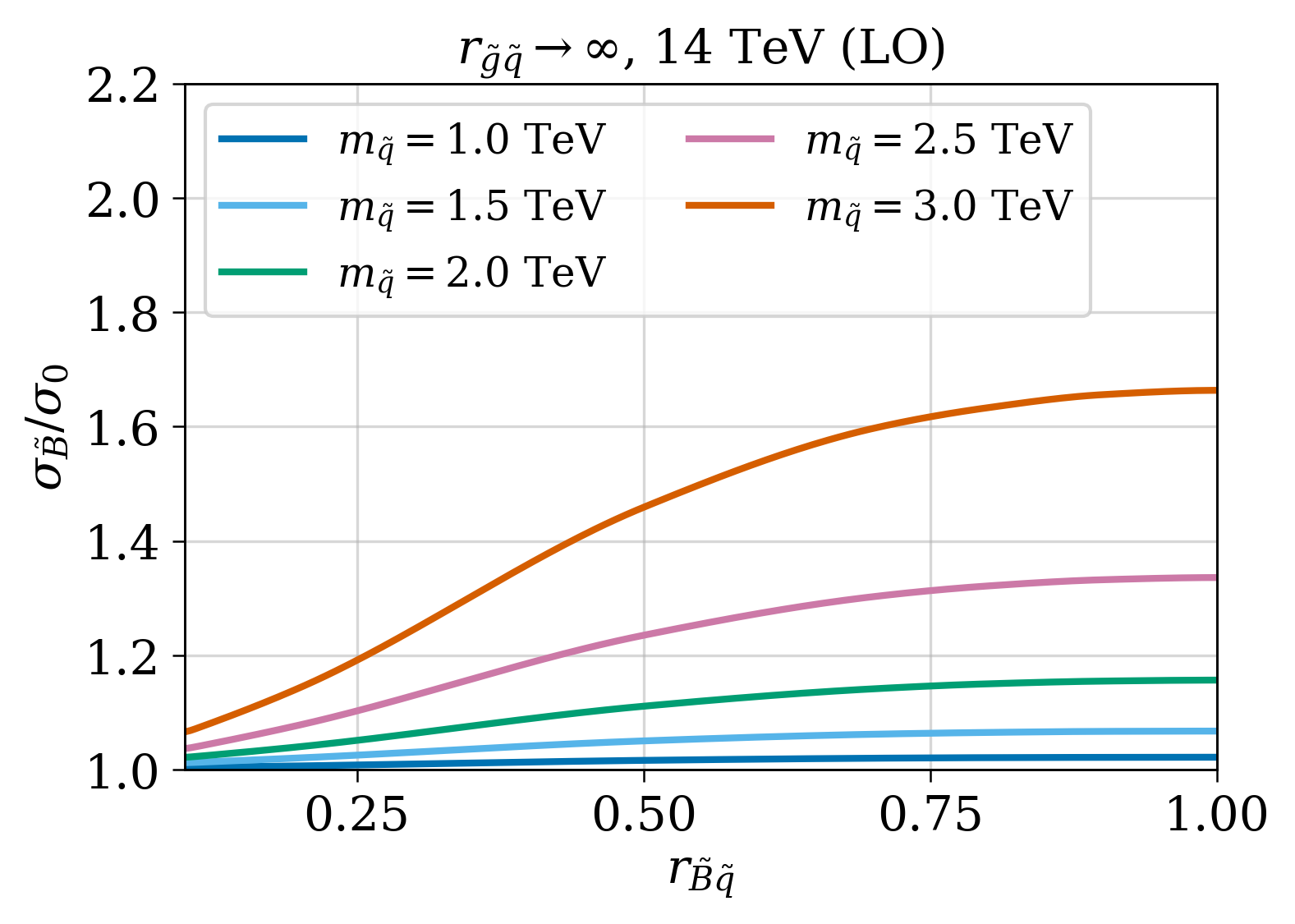}
\caption{Ratio of LO squark production cross sections at 14 TeV including only the Wino contribution ($\sigma_{\tilde{W}}$, left panel) or only the Bino contribution ($\sigma_{\tilde{B}}$, right panel) over the simplified-model prediction ($\sigma_0$).}
\label{fig:xsecRatioSimp}
\end{figure}

Our results for $\rap \to \infty$ are shown in \fig{fig:xsecRatioSimp}, for both the Wino and Bino contributions.  The Wino contribution can be a factor of a few larger than the reference cross section $\sigma_0$, showing that Winos can affect the predictions of the simplified model. The Bino contribution is more modest, but still slightly larger than $\sigma_0$, especially near the compressed region of small $\Delta=\msq -\mlsp$, where the experimental sensitivity is reduced.

\begin{figure}[h!]
\centering
\includegraphics[width=0.6\textwidth]{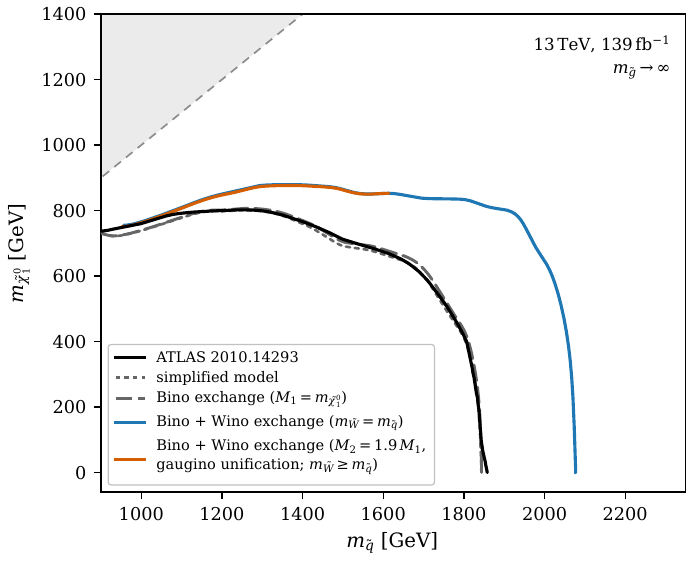}
\caption{Exclusion contours at $\sqrt{s}=13\TeV$ in the $\msq$--$\mlsp$ plane for the squark simplified model with the gluino decoupled: without EW gauginos (dotted), with $t$-channel Bino exchange for $m_{\tilde B}=\mlsp$ (dashed), and with Bino and Wino exchange for $m_{\tilde W}=\msq$ (blue) or for gaugino-mass unification, $m_{\tilde W}/m_{\tilde B}=1.9$ (orange), the latter shown only where $m_{\tilde W}\geq\msq$. The black line is the ATLAS limit~\cite{ATLAS:2020syg}. In all cases ${\rm BR}(\sq\to q\lsp)=100\%$.}
\label{fig:ewino}
\end{figure}

Figure~\ref{fig:ewino} shows how EW gauginos modify the LHC exclusion of the squark simplified model ($\rap\to\infty$). The Bino is the LSP, so its mass is fixed at every point on the plane, $m_{\tilde B}=\mlsp$, but its effect is marginal, changing the limit by less than $40\GeV$. Instead, the Wino mass is a free parameter. Its contribution is largest for $m_{\tilde W}\approx\msq$, which also keeps the decay $\sq_L\to q'\tilde W$ closed, so that the simplified-model assumption ${\rm BR}(\sq\to q\lsp)=100\%$ remains valid. In this case, the limit at $\mlsp=0$ moves from $1.84$ to $2.08\TeV$, corresponding to a cross-section enhancement by a factor $2.8$. With gaugino-mass unification, $m_{\tilde W}/m_{\tilde B}\simeq 1.9$, the Wino is heavier than the squarks only for $\mlsp\geq\msq/1.9$. In that region, the exclusion practically coincides with the $m_{\tilde W}=\msq$ case, since the Wino contribution depends only mildly on $m_{\tilde W}$ near its maximum. For lighter Winos, the squark decays are modified, and our analysis does not apply.

\section{The mirror question: is the gluino simplified model sound?}
\label{sec:gluino}

After having demonstrated that the squark simplified model fails to properly represent any realistic situation in low-energy supersymmetry, it is fair to ask whether the same criticism applies to the \emph{gluino} simplified
model, in which squarks are decoupled.

\renewcommand{\arraystretch}{1.5}
\begin{table}[t]
\centering
\begin{tabular}{|c|c|c||c|}
\hline
 {\small Parton } &  {\small Parton }&  {\small Threshold} & {\small $pp$ } \\[-3mm]
  {\small  process} &  {\small cross section}&  {\small behavior} & {\small cross section} \\
\hline
\hline
{\normalsize $gg\to \gl\gl$} & 
{\normalsize $\frac{27\pi\alpha_{s}^{2}\,\beta}{64\,\mgl^2}\,f_{\gl}(\beta)$} &
S-wave& {\normalsize ${\bar\sigma}_{0g}$}\\
& {\footnotesize $f_{\gl}(\beta)=\frac{1-\beta^2}{3}\big( \frac{17\beta^2-33}{3}+\frac{7-2\beta^2-\beta^4}{\beta}\, \ln \frac{1+\beta}{1-\beta}\big)$} & &\\
& {\footnotesize $=1+\frac{10\,\beta^2}{9}+\mathcal{O}(\beta^4)$} & &
\\[1mm]
\hline
{\normalsize $q\bar q\to \gl\gl$} & 
{\normalsize $\frac{\pi\alpha_{s}^{2}\,\beta}{3\,\mgl^2}\,\big( 1-\frac{4\beta^2}{3}+\frac{\beta^4}{3}\big)$} &
S-wave& {\normalsize ${\bar\sigma}_{0q}$}\\[1mm]
& 
{\normalsize $-\frac{4\pi\alpha_{s}^{2}\,\beta}{3\,\mgl^2}\,\big( 1-\frac{\beta^2}{3}\big)/\rapinv^2$} &
S-wave& {\normalsize $-{\bar\sigma}_{1}/\rapinv^2$}\\[1mm]
& 
{\normalsize $\frac{8\pi\alpha_{s}^{2}\,\beta}{3\,\mgl^2}\,\big( \frac{1+41\beta^2/54-\beta^4/6}{1-\beta^2}\big)/\rapinv^4$} &
S-wave& {\normalsize ${\bar\sigma}_{2}/\rapinv^4$}\\[2mm]
\hline
\end{tabular}
\renewcommand{\arraystretch}{1}
\caption{The gluino production channels. The squark-independent
``simplified" cross section is given by ${\bar\sigma}_{0}={\bar\sigma}_{0g}+{\bar\sigma}_{0q}$ and
${\bar\sigma}_{1,2}$ are the squark-mediated contributions discarded by the simplified model. In the large-$\rapinv$ limit, the cross-section coefficients ${\bar\sigma}_{i}$ are independent of $\rapinv$. The third column gives the threshold
behavior of each channel.}
\label{tab:crog}
\end{table}

\begin{figure}[h!]
\centering
\includegraphics[width=\textwidth]{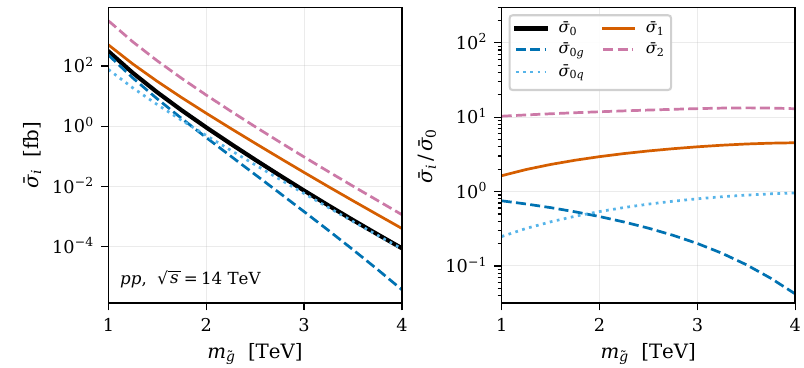}
\caption{Coefficients of the gluino-pair cross section against the gluino mass,
at $14\TeV$. Left panel: the four coefficients of \eq{segno} and the gluino simplified model cross section ${\bar \sigma}_{0}\equiv{\bar \sigma}_{0g}+{\bar \sigma}_{0q}$ (black curve). Right panel: cross-section coefficients in units of the reference value ${\bar \sigma}_{0}$.
The coefficient ${\bar \sigma}_1$ is plotted with the
sign convention of \eq{segno} and leads to a destructive interference.
}
\label{fig:gocoeff}
\end{figure}

As we have done for squarks, we begin by anatomizing the gluino production cross section, and the results are shown in \tab{tab:crog}. In the limit of decoupled squarks, gluinos can be produced via gluon fusion or quark-antiquark annihilation. 
The cross section for $\rapinv \to \infty$, where 
$$\rapinv =\msq/\mgl =1/\rap\,,$$
is governed by pure gauge interactions.
Both processes can occur in S-wave, as can be understood by taking into account color indices and Majorana properties. 

Squark $t$-channel exchange gives an additional contribution beyond the simplified model and the total production cross section can be decomposed as 
\beq
\sigma(pp\to \gl \gl)={\bar \sigma}_{0} -{{\bar \sigma}_1}/{\rapinv^2}+{{\bar \sigma}_2}/{\rapinv^4} \, ,~~~~{\bar \sigma}_{0}\equiv{\bar \sigma}_{0g}+{\bar \sigma}_{0q} \, ,
\label{segno}
\eeq
where ${\bar \sigma}_{0}$ is the $pp$ cross section corresponding to the gluino simplified model.
In the large-$\rapinv$ limit, the cross-section coefficients ${\bar \sigma}_{1,2}$ depend only on $\mgl$, and the dependence on the squark mass is fully accounted for by the explicit powers of $\rapinv$ in \eq{segno}. The values of the cross-section coefficients, as functions of the gluino mass, are shown in \fig{fig:gocoeff}.

In the case of squark production, the gluino mass insertion in the $t$-channel propagator is responsible for the $1/\rap^2$ contribution to the cross section, which determines the slow decoupling in \eq{forsigi}.
For gluino production, the squark exchanged in the $t$-channel is a scalar particle and therefore cannot benefit from the mass insertion. Nonetheless, $1/\rapinv^2$ contributions to the cross section exist because of the interference between $t$-channel squark exchange and the amplitude for $q\bar q\to g^*\to \gl \gl$. The interference is destructive: note the minus sign in the definition of ${\bar \sigma}_1$ in \eq{segno}. The interference effect of ${\bar \sigma}_1$ gives the leading departure from the simplified model, since the effect of ${\bar \sigma}_2$ is suppressed by $1/\rapinv^4$.

\begin{figure}[h!]
\centering
\includegraphics[width=\textwidth]{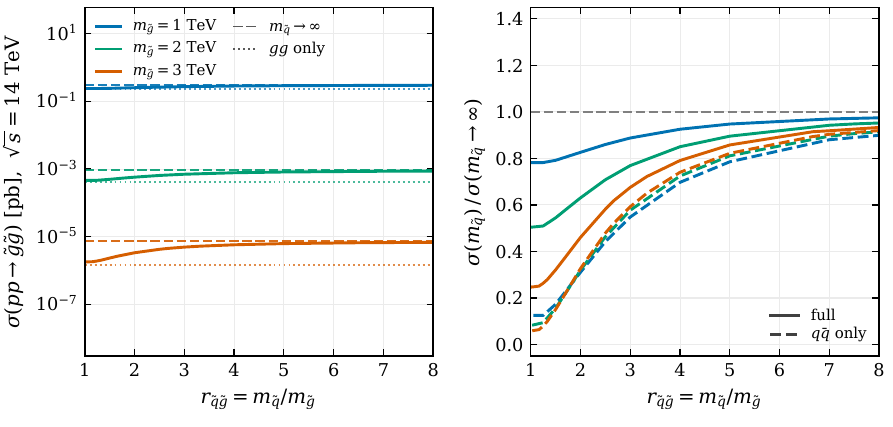}
\caption{Gluino pair production at $14\TeV$ (LO) versus $\rapinv=\msq/\mgl$, for
$\mgl=1,2,3\TeV$ (color). Left panel: $\sigma(pp\to\gl\gl)$; dotted is the
squark-independent $gg$ contribution, dashed the $\msq\to\infty$ (simplified-model)
value. Right panel: the same rate normalized to the ``simplified" value, for the full
process (solid) and for the $q\bar q$ channel alone (dashed), which carries the
destructive $s$-gluon $\times$ $t/u$-squark interference and dips near
$\msq\approx\mgl$.}
\label{fig:gluino}
\end{figure}

In \fig{fig:gluino}, we show the gluino production cross section and compare it with the simplified model prediction. As expected, finite squark masses reduce the gluino production cross section. The effect is significant for small $\rapinv$, but not as spectacular as in the case of squark production.
The reduction in cross section
grows with $\mgl$, partly due to the increase of the valence quark parton luminosity with $\shat$. Indeed, the squark-independent $gg$ contribution is
$75\%$ of the rate at $\mgl=1$~TeV and $20\%$ at 3~TeV. Moreover, the cross-section reduction dies out fairly quickly as $\rapinv$ grows. We find that, for $\rapinv=2$, the simplified model ($\rapinv \to \infty$) overestimates the LHC limit on the gluino mass by an insignificant 2\%.

In conclusion, we find that, unlike the case for squarks, the gluino simplified model gives an adequate description of realistic situations. This holds for three reasons.
\begin{itemize}
\item{There are no theoretical arguments that exclude large $\rapinv$ (unlike the case of $\rap$). Therefore, the approach to the simplified limit can correspond to realistic models.}
\item{The approach to the simplified limit occurs much faster than in the case of squarks. A slowly-decoupling contribution $1/\rapinv^2$ exists, but it comes from an interference effect, which is relatively modest in size.}
\item{The contribution from squark exchange does not have any particular enhancement factor with respect to the simplified model contribution (unlike the case of squark production).}
\end{itemize}
Sizable differences with the simplified model happen when $\rapinv \approx 1$. But, of course, the simplified model is not designed to describe situations in which gluinos and quarks are simultaneously produced at colliders.

\section{Conclusions}
\label{sec:conc}

We have demonstrated that low-energy supersymmetric models typically cannot lead to gluinos much heavier than squarks. This result is not contingent on any specific model; rather, it is a general feature of supersymmetric theories. Indeed, any initial mass hierarchy with gluinos heavier than squarks is destabilized at the quantum level and rapidly erased by radiative corrections.

In a large variety of supersymmetric models with high or moderately low mediation scale, the ratio $\rap =\mgl/\msq$ is typically less than 2. It can exceed 2 only when the mediation scale becomes very low and approaches the limit of non-perturbative direct mediation. Even in this extreme case, it is difficult to imagine sensible supersymmetric models in which $\rap$ is as large as 3. Our conclusion is that any realistic model of low-energy supersymmetry with Majorana gluinos satisfies the constraint $\rap \lsim 3$.

The squark simplified model often adopted in collider analyses corresponds to $\rap \to \infty$, and therefore misrepresents typical models of low-energy supersymmetry. This has important consequences for collider searches since the squark production cross section at hadron colliders is highly sensitive to $\rap$. For example, for $\msq =3\TeV$ and $\rap =2$, the simplified model underestimates the production rate at the LHC by more than a factor of 60. Therefore, adopting the squark simplified-model cross section is not merely a way of deriving more conservative lower bounds on $\msq$. Instead, it is a way of conveying misleading information regarding the status of low-energy supersymmetry. When applied to the LHC, the simplified-model limits suggest that certain ranges of $\msq$ are allowed, even though they are actually already excluded by current data in any realistic model. When applied to future colliders, the simplified-model cross section significantly underestimates their potential for discovering squarks.

The impact of $\rap$ in current limits to the squark mass and future searches is truly remarkable. The current LHC limit on $\msq$, based on the simplified model, is in the range 1.7--1.8~TeV. When realistic values of $\rap$ are considered, the current LHC limit is found to be in the range 2.6--3.0~TeV. While the HL-LHC reach on $\msq$ is predicted to be about 2.4~TeV on the basis of the simplified model, we find that the HL-LHC can realistically explore squarks up to 3.2--3.6~TeV. The commonly-quoted FCC-hh reach on squarks is 13~TeV. Instead, we find that masses up to 18--20~TeV are within reach, when realistic values of $\rap$ are considered.

Our recommendation is to discontinue the use of the squark simplified model in collider analyses for two main reasons discussed in our paper: {\it (i)} it does not represent any realistic low-energy supersymmetric model; {\it (ii)} it leads to misleading results when extracting lower bounds on $\msq$ at the LHC or when predicting the squark mass reach at future colliders. 

We recommend replacing the squark simplified model based on the parameters $(\msq,\,\mlsp)$ with a model described by the three parameters $(\msq,\,\mlsp,\,\rap \lsim 3)$. As long as the upper bound on $\rap$ is implemented, this choice cures the shortcomings {\it (i)} and {\it (ii)} of the squark simplified model. We emphasize that our proposal is not merely a continuous deformation of the squark simplified model, obtained by adding a third parameter. The class of theories described by the $(\msq,\,\mlsp,\,\rap \lsim 3)$ parametrization is disjoint from the class of simplified models parametrized by $(\msq,\,\mlsp)$, and there is no overlap between the predictions of these two classes. A key observation of this paper is that squarks in realistic low-energy supersymmetric scenarios populate the $(\msq,\,\mlsp,\,\rap \lsim 3)$ class of models, but leave empty the $(\msq,\,\mlsp)$ class of simplified models.

\subsubsection*{Acknowledgements}
 
KA would like to thank the organizers of the IV Joint ICTP-SAIFR/ICTP-Trieste School on Particle Physics and the Invisibles26 workshop for providing the atmosphere for stimulating discussions related to this work.
We thank Abhishek Banerjee, Alberto 
Belloni, Beate Heinemann, Zachary Marshall, Riccardo Rattazzi, Raman Sundrum and Felix Yu
for useful discussions and comments.
The work of KA is supported by 
NSF grant PHY-2609893 and the Maryland Center for Fundamental Physics.
AL is supported by FAPESP under grants no.~2018/25225-9 and 2021/01089-1 and CNPq grant no.~300278/2025-0.
RF and XW are supported in part by the European Union - Next
Generation EU   Missione 4 - Componente 2 - Investimento 1.1 
 CUP: I53D23000950006 through the MUR PRIN2022 Grant n.202289JEW4.
The work of XW is supported by the R\&D\&i project PID2023-146686NB-C31, funded by MICIU/AEI/10.13039/501100011033/ and by ERDF/EU, and by the Departament de Recerca i Universitats of the Generalitat de Catalunya through the Grup de Recerca 00649 (Codi: 2021 SGR 00649). IFAE is partially funded by the CERCA program of the Generalitat de Catalunya, and by the Spanish Ministry of Science and Innovation (MICINN) through the Spanish State Research Agency under the Severo Ochoa Centres of Excellence Programme 2025--2029 (CEX2024001442-S).
\appendix
\section{Recasting of the ATLAS search for squarks}
\subsection{Recasting materials\label{app:materials}}
In this appendix we collect the material from \cite{ATLAS:2020syg} that is necessary to reproduce our results. The exclusion curve in the $(\msq,\mlsp)$ plane for  eight degenerate squarks at $\rap \to \infty$   is given in Fig.~13 of ref.~\cite{https://doi.org/10.17182/hepdata.95664.v2/t46}.
The same search also considers the squark-gluino simplified model, but for fixed values of $\mlsp$. 
The exclusion curves for this  scenario can be found in refs.
~\cite{https://doi.org/10.17182/hepdata.95664.v2/t60,https://doi.org/10.17182/hepdata.95664.v2/t62,https://doi.org/10.17182/hepdata.95664.v2/t64}, which can be used to infer some information on the $\rap$ dependence.

To obtain limits for realistic $\rap$ values, we have used the upper limit map for the squark simplified-model cross section  provided as auxiliary material by ATLAS~\cite{https://doi.org/10.17182/hepdata.95664.v2/t65}. However, this map is limited to $\msq \leq 2.4$~TeV and was extrapolated for the higher $\msq$ values used in our results. The extrapolation is anchored to the seven points excluded by ATLAS in Fig.~17 of ref.~\cite{ATLAS:2020syg}, which, by definition of exclusion, provide measurements of $\sigma_{UL}$ at $\msq=2.5$--$3.0\TeV$.

\subsection{Cross-section calculation}
\label{app:cross-section}

The computation of $\sigmatot$ used by ATLAS~\cite{ATLAS:2020syg} includes approximate NNLO QCD corrections and resummation up to NNLL (dubbed as \nnllfast) with the
 parton distribution functions described by \textsc{PDF4LHC15} \cite{Butterworth:2015oua}. The cross section is obtained using \textsc{NNLL-fast~2.0}\cite{Beenakker:2016lwe,Beenakker:2024jwh}, which is capable of computing the rates for $\rap\to\infty$ and $\rap=1$ for some discrete predefined values of squark and gluino masses. 
The resulting cross sections are 10--40\% larger than NLO rates computed with \textsc{Prospino2}~\cite{Beenakker:1996ch}. In the $\sq\sq$ channel, which dominates the rate for large $m_{\sq}$ and $\rap<3$, the increase from NLO to \nnllfast\ is more moderate, around 10\%.
At high squark masses, the PDF set dependence for the dominant $\sq\sq$ channel is found to be larger than the rescaling from NLO to \nnllfast\ that we apply. However, the squark mass bound is subject only to a moderate uncertainty around few percent.\footnote{We stress that, for future squark searches at high mass, it will be crucial to test PDF fits against the contamination of new physics at high-$x$ as discussed in Refs.~\cite{Cole:2026eex,Harland-Lang:2025wvm,Azzi:2019yne,AbdulKhalek:2018rok,PDF4LHCWorkingGroup:2022cjn,Beenakker:2015rna,Beenakker:2024jwh}.}
As \textsc{Prospino2} can compute NLO accurate cross sections for generic squark and gluino mass, we use it for our cross-section calculation.  We then apply a $K$ factor to bring the cross section to   \nnllfast\ accuracy, computed using \textsc{NNLL-fast~2.0} (or the LHC SUSY Cross Section Working Group TWiki page \cite{TwikiNNLLfast}) for values of masses for which the computation of \nnllfast\ is publicly released. Outside of this discrete set of masses, we interpolate from neighboring points around the point of interest. 
For   $m_{\lsp}<995\text{ GeV}$ and $m_{\sq}<3000\text{ GeV}$, our interpolation of the public information from \cite{ATLAS:2020syg,hepdata.95664} is always an intra-polation. Only for heavier $\lsp$ and $\sq$ we need to extra-polate outside of the ranges covered by ATLAS public information. In particular, we remark that staying away from the  so-called ``compressed spectrum'' region,  in which the  squark-neutralino  mass degeneracy always requires a special treatment or a whole dedicated analysis, we can use public data to intra-polate, giving us confidence in the reliability of our results. 

\subsection{Kinematics and other caveats of the recast \label{app:caveats}}

The extraction of a limit on the squark mass requires the computation of observed cross-section bounds ($\sigma_{UL}$) for scenarios with finite $\rap$.
Although ATLAS has provided values for $\sigma_{UL}$ for the simplified model ($\rap \to \infty$) case, these do not necessarily apply to realistic models, where the $t$-channel gluino contribution is sizable.
Since the gluino contribution can, in principle, modify the kinematics of the signal, the acceptances and efficiencies for finite $\rap$ may significantly differ from the simplified model scenario.
We checked, however, that assuming similar efficiencies/acceptances is a rather accurate approximation.
Indeed, 
applying the SR2j-2800 selection verbatim from the published cut-flow, the two
acceptances for $\sq\sq$ and $\sq\sq^*$ at parton-level agree to $1$--$2\%$ in the range $\msq=1.6$--$3.2\TeV$,
$\rap=1$--$10$ and $\mlsp=0$--$1495\GeV$.

\textbf{Compressed region.} The contours in \fig{fig:reach} are not drawn below $\Delta=\msq-\mlsp=60\GeV$. For $60\GeV<\Delta\lsim200\GeV$, the most sensitive ATLAS selection is the compressed one, and the contours rely on an extrapolation of the published $\sigma_{UL}$ grid, which is sparse and irregular in this region, to squark masses beyond those covered by the grid. No finite-$\rap$ measurement by ATLAS is available to anchor this extrapolation, and the equality of the $\sq\sq$ and $\sq\sq^*$ acceptances has been checked only for the standard selection. In this region, the contours should therefore be regarded as indicative only.

\section{Projection to higher energies and luminosities \label{app:projections}}

The bound on the squark mass obtained at $\sqrt{s}=13\TeV$ with an integrated luminosity $\mathcal{L}=139\fb^{-1}$ can be read as a bound at the corresponding value of $\tau=4\msq^2/s$. It is set by the SM background at that value of $\tau$ and by the signal acceptance. The background cross section scales approximately as $\sigma_B\sim F(\tau)/s$, {\it i.e.}~it depends on the collider energy through the overall factor $1/s$, and its remaining $\tau$ dependence is nearly universal. For a search limited by statistical uncertainties, the upper limit scales as $\sigma_{UL}\propto\sqrt{\sigma_B/\mathcal{L}}$. The limit at a collider with center-of-mass energy $\sqrt{s'}$ and integrated luminosity $\mathcal{L}'$ can then be estimated from the ATLAS result as
\begin{equation}
\sigma_{UL}(\msq',\Delta';\,\sqrt{s'},\mathcal{L}')\;=\;\kappa\,
\sigma_{UL}^{\textrm{ATLAS}}(\msq,\Delta;\,\sqrt{s},\mathcal{L})\,,
\quad
\kappa\;=\;\rho\,\sqrt{\frac{\mathcal{L}}{\mathcal{L}'}}\;\sqrt{\Phi\,\frac{{\alpha'_s}^2}{\alpha_s^2}}\,.
\label{eq:fccscale}
\end{equation}
The masses are rescaled as $\msq'=\msq/\rho$ and $\Delta'=\Delta/\rho$, with $\rho=\sqrt{s}/\sqrt{s'}$, and $\alpha_s$ and $\alpha'_s$ are the strong couplings at $\msq$ and $\msq'$. The factor $\Phi=L_{pp}(\tau,\msq')/L_{pp}(\tau,\msq)$ is the ratio of parton luminosities at the two factorization scales; both are evaluated at the same $\tau$ because, once all dimensionful selection thresholds of the analysis of ref.~\cite{ATLAS:2020syg} are scaled by $1/\rho$, the projected bound lies at approximately the same $\tau$.

The squark mass bound then follows from the cross-section bound of \eq{eq:fccscale} and the squark production rate, computed with the method of \app{app:cross-section}.

\bibliographystyle{jhep}
\bibliography{bibfile.bib}

\end{document}